\documentclass[%
 aip,
pop,
showkeys,
amsmath,
amssymb,
reprint,
onecolumn,
]{revtex4-2}

\usepackage{amsmath}
\usepackage{amsthm}
\usepackage{amssymb}
\usepackage{color}
\usepackage{xcolor}
\usepackage{cases}
\usepackage{enumitem}             
\usepackage{mathtools}
\usepackage{siunitx}
\usepackage[OMLmathsfit]{isomath}  
\usepackage{graphicx}
\usepackage[T1]{fontenc}
\usepackage{lmodern}

\usepackage[cmintegrals]{newpxmath}  
\usepackage{bm}                  
          
\begin{document}

%
%
\title{A Multiscale Eulerian Vlasov-Rosenbluth-Fokker-Planck Algorithm for Thermonuclear Burning Plasmas}

\author{Benjamin L. Reichelt}
\affiliation{Theoretical Division, Los Alamos National Laboratory, Los Alamos, NM 87545, USA}
\affiliation{Plasma Science and Fusion Center, Massachusetts Institute of Technology, 77 Massachusetts Ave., NW17, Cambridge, MA 02139, USA}

\author{William T. Taitano}
\affiliation{Theoretical Division, Los Alamos National Laboratory, Los Alamos, NM 87545, USA}

\author{Brett D. Keenan}
\affiliation{Computational Physics Division, Los Alamos National Laboratory, Los Alamos, NM 87545, USA}

\author{Luis Chac{\'o}n}
\affiliation{Theoretical Division, Los Alamos National Laboratory, Los Alamos, NM 87545, USA}

\author{Andrei N. Simakov}
\affiliation{Theoretical Design Division, Los Alamos National Laboratory, Los Alamos, NM 87545, USA}

\author{Steven E. Anderson}
\affiliation{Computational Physics Division, Los Alamos National Laboratory, Los Alamos, NM 87545, USA}

\author{Hans R. Hammer}
\affiliation{Theoretical Division, Los Alamos National Laboratory, Los Alamos, NM 87545, USA}

%
%
%
%
%
%

\date{\today}

\begin{abstract}
Accurate treatment of energetic fusion byproducts in laboratory plasmas often requires a kinetic description, owing to their large birth kinetic energy and long mean free paths compared with the characteristic system scale lengths. For example, alpha particles produced by deuterium-tritium fusion reactions are born at high energies (\qty{3.5}{MeV}) and predominantly slow down through interactions with electrons traveling at comparable speeds. As an alpha particle slows, its distribution collapses near the background ion-thermal speed, forming a sharp structure in velocity space. Such sharp features pose numerical challenges in grid-based Eulerian methods: capturing the full alpha-particle energies demands a large velocity domain, while resolving the near-thermal region requires a sufficiently fine mesh. Inspired by the work of Peigney \textit{et al.}~[\textit{J.\ Comput.\ Phys.} \textbf{278} (2014)], we present a two-grid approach that splits the alpha-particle distribution into energetic (suprathermal) and ash (thermal) components. A Gaussian-based sink term transfers particles from the energetic population to the ash population as they slow to the thermal regime, and a conservative projection scheme ensures that mass, momentum, and energy of the slowing energetic alpha population are conserved when it is remapped onto the ash grid. Unlike the formulation of Peigney \textit{et al.}, our method does not require a strict asymptotic separation of velocity scales, which can, in principle, be arbitrary. We demonstrate the robustness of this approach on challenging multiscale problems, including a surrogate for an igniting inertial confinement fusion capsule.
\end{abstract}

\maketitle

%
%
\section{Introduction}
\label{sec:introduction}
The Vlasov-Fokker-Planck (VFP) equation describes the dynamical evolution of the plasma particle distribution function (PDF), \(f(t,\mathbf{x},\mathbf{v})\), in phase space, where \(t \in \mathbb{R}_+\) is time, \(\mathbf{x} \in \mathbb{R}^3\) denotes particle position, and \(\mathbf{v} \in \mathbb{R}^3\) denotes particle velocity. The equation accounts for transport of charged particles under electromagnetic fields and includes binary collisions mediated by cumulative small-angle Coulomb scattering. When coupled to Maxwell's equations, the VFP equation is recognized as a high-fidelity model for weakly coupled plasmas, valid across all collisionality regimes.

In inertial confinement fusion (ICF) systems, thermonuclear fusion reactions and their impact on plasma dynamics and energetics must be modeled to accurately describe the burning process. Physically, particles produced by these reactions (e.g., \(\alpha\)-particles from deuterium-tritium (DT) fusion at 3.5\,MeV) preferentially slow down through interactions with electrons moving at comparable thermal speeds, \(v_{\text{th},e} = \sqrt{2T_e/m_e}\), where \(T_e\) and \(m_e\) are the electron temperature and mass, respectively. As an \(\alpha\)-particle slows down, its distribution stretches from the birth velocity \(v_{\alpha}\) (corresponding to birth energy \(\sim 3.5\)\,MeV) to thermal speeds corresponding to fuel temperatures around 5--10\,keV, producing a sharp structure in velocity space. Once the \(\alpha\)-particles reach thermal energies, they become part of the so-called \emph{ash} population, with \(v_{\text{th},A} \ll v_{\alpha}\). In terms of the characteristic speed ratio \(v_{\alpha}/v_{\text{th,A}}\), this separation can be as large as a factor of 20--30 in the fuel hot spot. Numerically, the velocity domain must span up to \(v_{\alpha}\), and the grid must be fine enough near \(v_{\text{th},A}\) to capture the near-singular feature, leading to a computational cost scaling approximately like \(\bigl(v_{\alpha}/v_{\text{th},A}\bigr)^d\), where \(d\) is the dimensionality in velocity space. This challenge is exacerbated in modern ICF experiments, such as the recent hybrid-E designs that achieved fusion ignition\cite{abu_shawareb_2022_prl_ignition}. The thermonuclear fuel typically has a central vapor region (the ``hot spot'') at 5--10 keV, surrounded by cooler cryogenic DT ice at 10--50 eV. Consequently, the velocity-scale separation between fusion-born \(\alpha\)-particles and the cold regions can reach factors of 260--600, making a uniform velocity mesh impractical.

One natural strategy to address such disparate scales is adaptive mesh refinement (AMR). However, developing a practical AMR framework for the VFP equation poses significant numerical and computational challenges. Specifically, AMR methods must simultaneously deal with stiff anisotropic diffusion and advection in velocity space, maintain distinct mesh hierarchies for each plasma species \textit{and} spatial location, conserve critical physical invariants (mass, momentum, and energy), and correctly recover the hydrodynamic limit under strongly collisional conditions. Similarly, general curvilinear adaptive mesh techniques, which locally deform meshes to follow evolving thermal structures, introduce considerable algorithmic complexity, particularly in ensuring stability, accuracy, and conservation of physical quantities on dynamically adapting grids. Such complexity not only complicates numerical implementation but also may significantly degrade computational efficiency in realistic, multiscale fusion scenarios.

In contrast, the two-grid approach pioneered by Peigney \textit{et al.}\cite{peigney2014alpha} leverages a simpler yet powerful idea: employing two separate uniform grids for energetic $\alpha$-particles and the thermal \emph{ash} population, coupled through physically motivated numerical source and sink terms. This strategy inherently avoids the complexity and overhead associated with AMR or dynamically adapting general curvilinear meshes, significantly simplifying numerical implementation, reducing computational costs, and enhancing robustness. Successfully demonstrated in the FPion and FUSE codes\cite{larroche2003kinetic_omega, larroche2003kinetic_rygg, larroche2007diffusionexpliti, larroche2018omega}, this approach has effectively simulated multiscale capsule implosions at the National Ignition Facility (NIF)\cite{peigney2014alpha_physics, inglebert2014_neutron_diagnostics}. Nonetheless, Peigney's method relies critically on an asymptotic scale-separation assumption that may fail under realistic burning-plasma conditions, where fuel and ash temperatures reach tens of keV. Additionally, their approach lacks rigorous convergence studies, leaving open questions about consistency with the original VFP equation. Furthermore, the approach relies on a reformulation of the underlying VFP equations using a mixture of asymptotic analysis and ad-hoc modifications as well as on a Lorentz collision operator rather than the full Landau-Rosenbluth-Fokker-Planck model. Such approximations may lead to a loss of continuum detailed balance of momentum and energy. Finally, to satisfy the analytical asymptotic reduction procedure numerically, a hybrid coordinate was introduced whereby the $\alpha$ population was represented in spherical coordinates while the ash was represented in cylindrical coordinates. These specialized choices challenge the generalizability of the method and adoption by other codes that lack such capabilities. Despite these limitations, the conceptual simplicity, ease of implementation, and computational tractability of Peigney's two-grid method could make it an attractive alternative to more complex adaptive strategies, particularly for modeling dynamically heated fusion plasmas in ICF conditions.

In this study, we build upon Peigney's original idea to develop an Eulerian two-grid strategy that lifts the limitations described above. As before, we employ separate velocity-space grids for energetic \(\alpha\)-particles and thermal ash; however, we design the numerical \emph{source and sink} terms in a way that ensures consistency with the original VFP equation and strictly preserves the detailed balance of mass, momentum, and energy as energetic alphas are remapped onto the ash grid. Our approach uses a physics-based, reduced-order description of the \(\alpha\)-ash interaction to construct the sink term on the \(\alpha\)-particle grid; this term is then mapped to the ash grid in a way that preserves detailed balance in the continuum. Numerically, the resulting discretization introduces small truncation errors in conservation, which we eliminate through a projection step. We also develop a suite of numerical techniques to handle extreme cases involving dense, cold DT ice. This consistent and symmetry-preserving two-grid method allows us to capture multiscale slowing-down and thermal-structure formation without explicitly resolving such scales on the \(\alpha\)-particle grid or relying on strict scale separation for the formulation to be valid. Furthermore, the formulation employs the same (cylindrical) coordinates for both ash and $\alpha$ populations and does not require specialized treatment for a particular population. As a result, we obtain a robust framework for investigating kinetic effects in thermonuclear plasmas.

The remainder of this manuscript is organized as follows. In Section~\ref{sec:hybrid_fluid_electron_and_ion_fp_model}, we introduce the ion-VFP equation with a fusion-reaction source term, coupled to a fluid electron-energy equation. Because the main novelty concerns the interplay between energetic \(\alpha\)-particles and ash, our discussion focuses on a spatially homogeneous setting. In Section~\ref{sec:conservative_multiscale_2_grid_strategy}, we describe our principal contribution: the {\em consistent and conservative} two-grid strategy at the continuum level. The numerical implementation details are provided in Section~\ref{sec:numerical_implementation}. In Section~\ref{sec:numerical_results}, we demonstrate the algorithm's performance on problems with varying complexity, highlighting its ability to handle ignition-scale ICF scenarios. We conclude in Section~\ref{sec:conclusions}.
%
%
%
%
%
\section{Hybrid Fluid-Electron and Ion-Fokker-Planck Model with Thermonuclear Fusion Source}
\label{sec:hybrid_fluid_electron_and_ion_fp_model}
The spatially homogeneous Fokker-Planck equation for ion plasma species $\gamma$ in this study with a source is given as:
\begin{flalign}
    \label{eqn:vfp}
    &
    \partial_t f_{\gamma} = \sum_{\gamma'\in {\cal G}}C_{\gamma\gamma'} + C_{\gamma e} + S_{\gamma}.
    &
\end{flalign}
Here, $f_{\gamma}\left( {\bf v},t \right)$ is the distribution function, 
\begin{flalign}
    \label{eqn:ion_fp}
    &
    C_{\gamma\gamma'} = \Gamma_{\gamma\gamma'} \nabla_v \cdot \left[ {\cal D}_{\gamma'} \cdot \nabla_v f_{\gamma} - \frac{m_{\gamma}}{m_{\gamma'}} {\bf A}_{\gamma'} f_{\gamma} \right],
    &
\end{flalign}
is the complete Rosenbluth-Fokker-Planck collision operator for the ion species $\gamma$ colliding with species $\gamma'$, ${\cal G}$ is the set of ion species indices (${\cal G} = \left\{D,T,\alpha,A \right\}$ in this study corresponding to deuterium, tritium, $\alpha$ and ash, respectively), ${\cal D}_{\gamma'} \left[ f_{\gamma'}\right]\left({\bf v}\right) = \nabla_v \nabla_v G_{\gamma'}$ is the symmetric-positive-definite collisional diffusion tensor, ${\bf A}_{\gamma'} \left[f_{\gamma'}\right] = \nabla_v H_{\gamma'}$ is the collisional friction vector, $G_{\gamma'}\left[f_{\gamma'}\right] \left({\bf v}\right)$ and $H_{\gamma'}\left[f_{\gamma'}\right] \left({\bf v}\right)$ are the Rosenbluth potentials, obtained from the solutions to the coupled Poisson equations,
\begin{flalign}
    \label{eq:H_poisson_equation}
    &
    \nabla^2_v H_{\gamma'} = -8\pi f_{\gamma'}
    &
\end{flalign}
\begin{flalign}
    \label{eq:G_poisson_equation}
    &
    \nabla^2_v G_{\gamma'} = H_{\gamma'},
    &
\end{flalign}
$\Gamma_{\gamma\gamma'} = \frac{Z^2_{\gamma}Z^{2}_{\gamma'} e^4 \ln{\Lambda_{\gamma\gamma'}}}{8\pi \epsilon^2_{0} m^2_{\gamma}}$, $Z_{\gamma}$ is the charge number of the $\gamma$ species, $e$ is the elementary charge, $\epsilon_0$ is the permittivity constant of vacuum, $m_{\gamma'}$ is the mass of $\gamma'$ species, $\ln{\Lambda}_{\gamma\gamma'}$ is the Coulomb logarithm between species (which we take to be 10 for all interactions in this study), 
\begin{flalign}
    \label{eqn:ie_fp}
    &
    C_{\gamma e} = \nu_{\gamma e} \nabla_v \cdot \left[ 
    \frac{T_e}{m_{\gamma}} \nabla_v f_{\gamma} - \left({\bf u}_{\gamma} - {\bf v} \right)f_{\gamma}
    \right]
    &
\end{flalign}
is the ion-electron Lenard-Bernstein-Fokker-Planck operator\cite{hazeltine_plasma_confinement_1991}, which is the asymptotic model for describing the collisions of slow ions on Maxwellian electrons, i.e., $v_i \ll v_{th,e}$, 
\begin{flalign}
    \label{eqn:nu_ie}
    &
    \nu_{\gamma e} = \frac{\sqrt{2m_e} n_{e} Z^2_{\gamma} e^4 \ln \Lambda_{\gamma e}}{12 \pi^{3/2} m_\gamma \epsilon^2_0 T^{3/2}_e}
    &
\end{flalign}
is the ion-electron collision frequency, $T_e$ is the electron temperature, ${\bf u}_{\gamma}$ is the drift velocity, and $S_{\gamma}$ is the source/sink for species $\gamma$. The electron number density and velocity are determined by quasi-neutrality, \(n_e = -q^{-1}_e \sum_{\gamma \in {\cal G}} q_{\gamma} n_{\gamma}\), and ambipolarity, \({\bf u}_e = - q^{-1}_e n^{-1}_{e} \sum_{\gamma \in {\cal G}} q_{\gamma} n_{\gamma} {\bf u}_{\gamma} \) while the internal fluid electron energy equation is evolved by:
\begin{flalign}
    \label{eqn:fluid_e}
    &
    \frac{3}{2}\frac{\partial }{\partial t}\left(n_e T_e\right) = 3\sum_{\gamma \in {\cal G}} n_e \nu_{e\gamma} \left(T_{\gamma} - T_{e} \right).
    &
\end{flalign}
For notational convenience in the electron energy equation we define the electron energy-exchange frequency $\nu_{e\gamma}$ through $n_e \nu_{e\gamma} \equiv n_{\gamma}\nu_{\gamma e}$.

In the limit where no source/sink is present, the system supports a rigorous conservation theorem for mass, momentum, and energy; and the numerical preservation strategy of these properties used in this study are found in Ref.~\onlinecite{taitano_cpc_2021_ifp_code}.

The isotropic D+T $\rightarrow \alpha$ reaction source is given as:
\begin{flalign}
    \label{eqn:fusion_source}
    &
    S_{\alpha}\left({\bf v}\right) = \frac{R_{DT}}{4\pi v^2}\delta\left(v - v_{\alpha} \right),
    &
\end{flalign}
where we use the Maxwellian-averaged form 
\begin{flalign}
    \label{eqn:R_dt}
    &
    R_{DT} = n_{D}n_{T} \left< \sigma v \right>
    &
\end{flalign}
for the DT reactivity instead of evaluating expensive five-dimensional reactivity integrals. Here, $v = \left|{\bf v}\right|$, $v_{\alpha}$ is the speed at which the fusion $\alpha$ particle is born, $n_{D}$ and $n_{T}$ are the D and T number densities, and $\left< \sigma v \right> \left( T_{DT}; \boldsymbol{\xi} \right)$ is the Maxwellian averaged DT fusion reactivity, $T_{DT} = \frac{\left(n_D T_D + n_T T_T \right)}{n_D + n_T}$ is the average DT temperature, and $\boldsymbol{\xi}$ is the vector of parameters for the cross-section obtained in this study from the Bosch-Hale parameterization\cite{bosch_hale_1992_nucl_fusion_reactivity_legend_paper}; the parameterization for the DT-$\alpha$ reaction is provided in Appendix~\ref{app:bosch_hale_parameterization} for completeness.

%
%
\section{A Conservative Multiscale Two-Grid Strategy for Fusion-Born Energetic Particles}
\label{sec:conservative_multiscale_2_grid_strategy}
\begin{figure}[h!]
    \centering
    \includegraphics[scale=0.6]{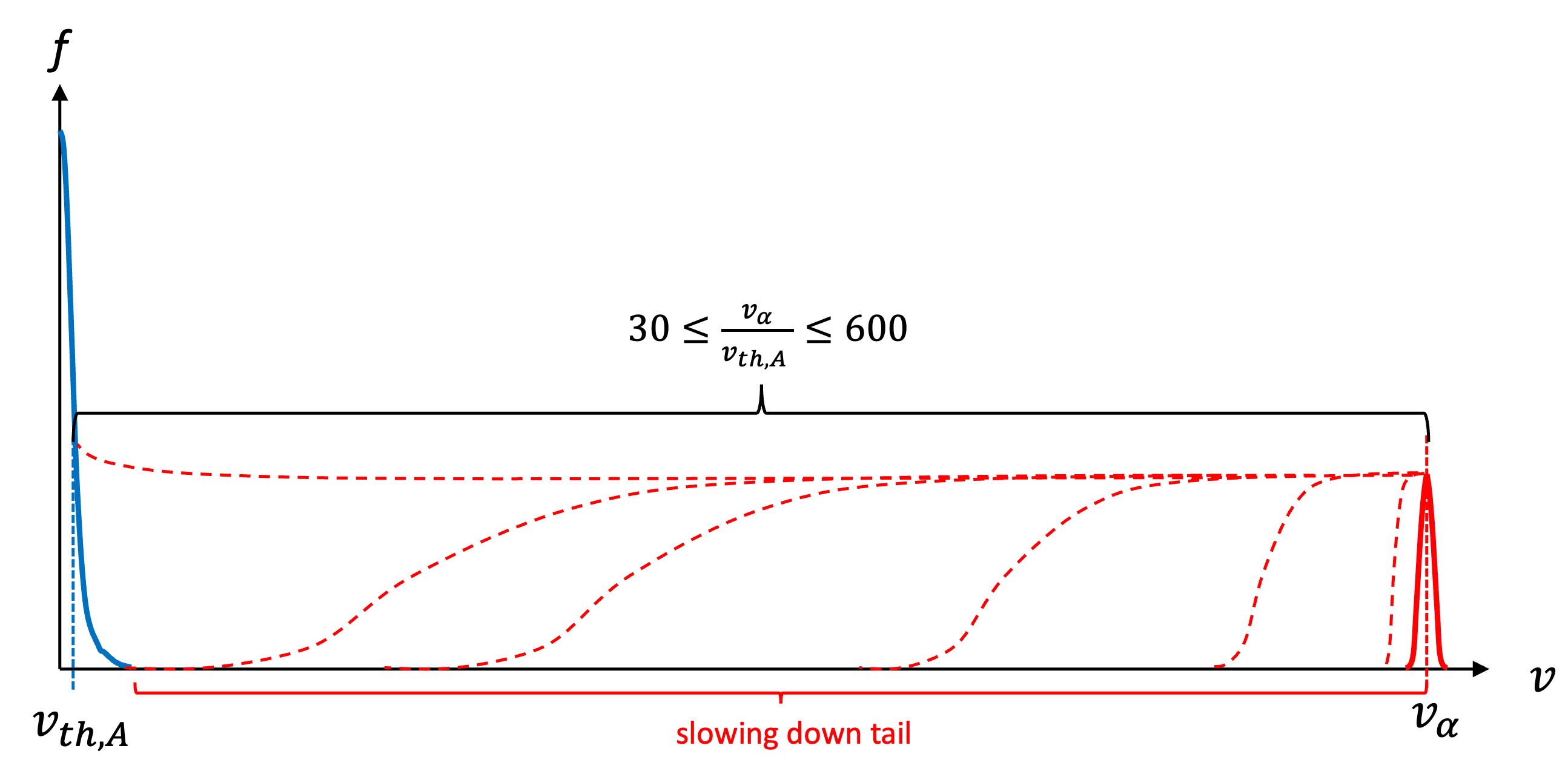}
    \caption{Illustration of velocity-space scale separation between the fusion $\alpha$ particles (red solid line) and the ash population (blue solid line), and the slowing down tail formation process (red dashed lines). The ash population is formed from the slowing-down tail population that eventually thermalizes with the $DT$ fuel population.}
    \label{fig:multiscale_illustration}
\end{figure}

In grid-based approaches, the reactivity source terms present unique multiscale challenges. We consider the \(DT-\alpha\) reactions, where the reacting \(DT\) ions typically possess energies in the range of \(5\) to \(10\,\text{keV}\), while the \(\alpha\) particles are produced with energies of approximately \(3.5\,\text{MeV}\). Upon birth, the \(\alpha\) particles preferentially collide with electrons, whose thermal speed is given by $v_{th,e} = \sqrt{\frac{2T_e}{m_e}},$ and decelerate accordingly. Once they slow down to a critical velocity, $v_c = \left( \frac{3\sqrt{\pi} \, T_e^{3/2}}{\sqrt{2m_e}\, n_e} \sum_{\gamma \in {\mathcal{G}^{\backslash \alpha}}} \frac{Z_\gamma^2\, n_\gamma}{m_\gamma} \right)^{1/3}$, they primarily thermalize with the \(DT\) ions, thereby forming the \emph{ash} population. These two distinct populations of \(^4\text{He}\) species---the fast \(\alpha\) particles and the slower ash---are separated by roughly a factor of \(30\) in velocity. Consequently, naive grid-based methods must employ a sufficiently large velocity-space domain to capture the high-energy \(\alpha\) particles while maintaining a fine grid resolution to accurately resolve the ash population (see Figure~\ref{fig:multiscale_illustration}).
When considering a cryogenic, high-density DT ice layer with electron temperatures in the range $10\mathrm{eV} \le T_e \le 50\mathrm{eV}$ (as shown in Sec.~\ref{subsec:1d2v_layered_spherical_implosion}), the scale separation can reach up to approximately 600, further underscoring the limitations of a naive grid-based approach.

\begin{figure}[h!]
    \centering
    \includegraphics[width=0.7\linewidth]{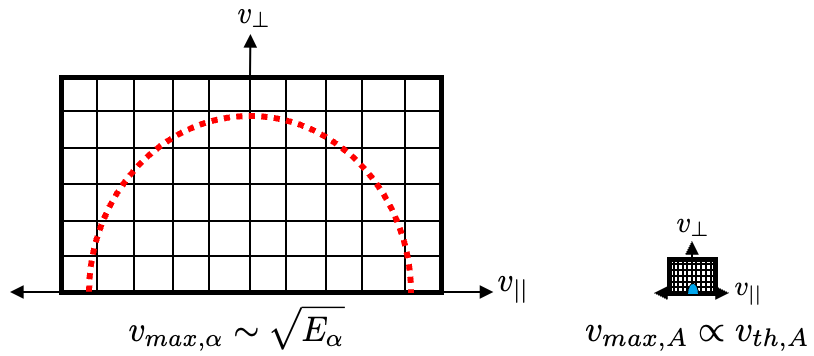}
    \caption{Illustration of individual grids (not to scale) for energetic species such as $\alpha$-particles (left) and slow ash population (right). The dashed-red semi-circle depicts the isotropic source of $\alpha$ particles born at speeds of $v_{\alpha}$ and energy $E_{\alpha}$ while the blue filled circle on the figure on the right depicts the thermal population of the ash species.}
    \label{fig:2_grid_approach}
\end{figure}

We propose a multiscale approach that models the fast, energetic population and the slower, thermal population on separate grids. In this study, we adopt an azimuthally symmetric, transformed two-dimensional cylindrical velocity-space coordinate system, denoted as $\widehat{\bf w} = \left\{\widehat{w}_{\|}, \widehat{w}_{\perp} \right\}$, where $\widehat{w}_{\|}$ and $\widehat{w}_{\perp}$ represent the transformed parallel and perpendicular velocity coordinates, respectively. The mapping to the original velocity coordinates is given by:
\begin{flalign}
    \label{eq:map_to_original_coordinate}
    &
    {\bf v} = v^*_{\gamma} \widehat{\bf w} + {\bf u}^*_{\gamma},
    &
\end{flalign}
where $v^*_{\gamma}$ is a scaling factor, and ${\bf u}^*_{\gamma} = \left\{ u^*_{\|,\gamma} , 0 \right\}$ represents a shift velocity for species $\gamma$. Thus, prescribing species-dependent values for $v^*_{\gamma}$ and ${\bf u}^*_{\gamma}$ effectively defines individual grids for each species. A graphical illustration of this two-grid approach is presented in Figure~\ref{fig:2_grid_approach} for fusion reaction-born $\alpha$ particles and ash species.
In general, the velocity space transformation metrics can be a function of time {\it and} configuration space (for spatially heterogeneous cases), which introduces a set of inertial terms in the transformed equations. The detailed discussion of such modifications to the equations are outside the scope of this work and we refer curious readers to Ref.~\onlinecite{taitano_cpc_2021_ifp_code} for a detailed theoretical and numerical treatment for the general scenario. 

\begin{figure}[h!]
    \centering
    \includegraphics[width=0.4\linewidth]{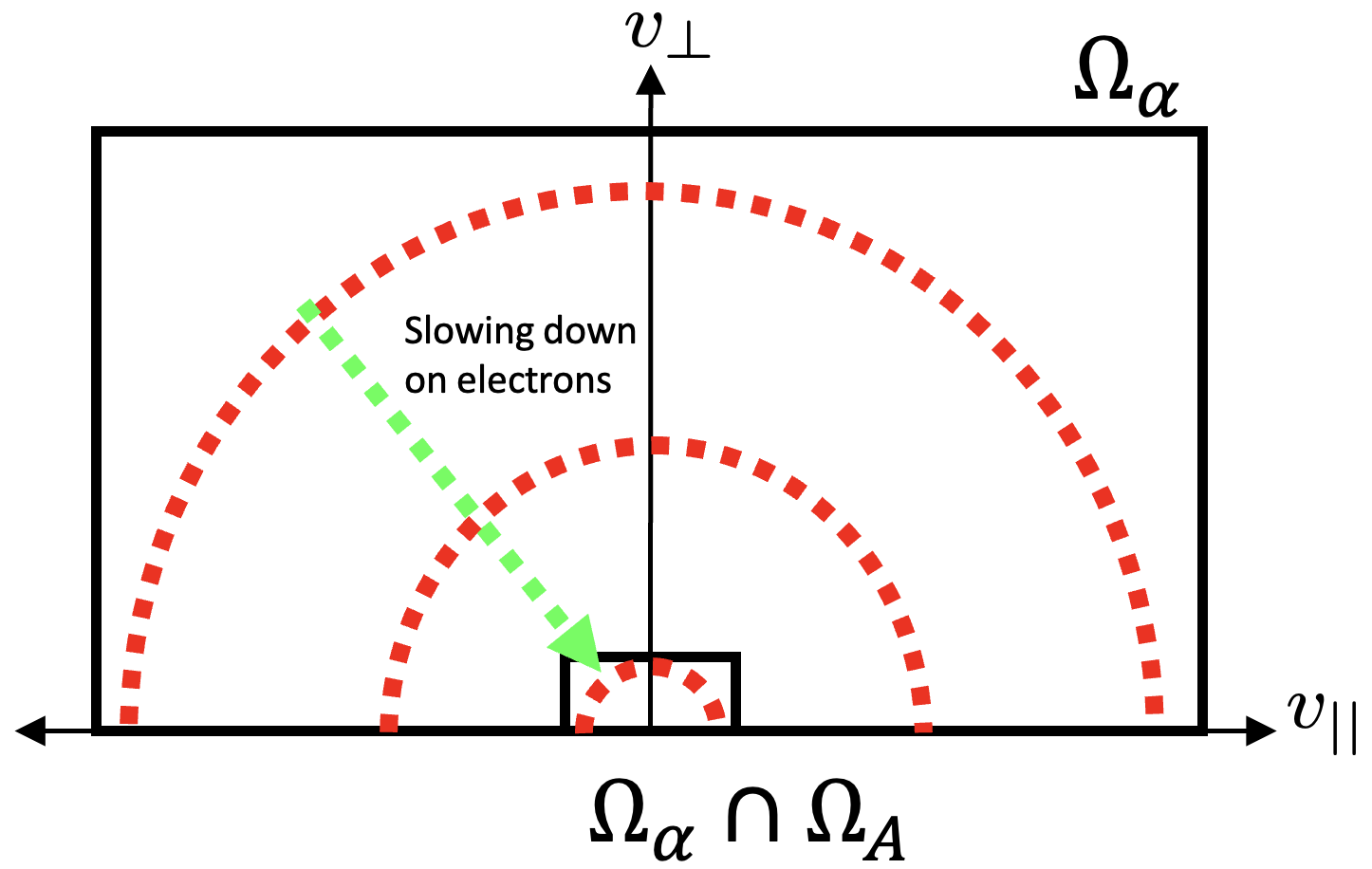}
    \caption{The slowing down process of energetic alphas. The dotted red lines represent the energetic particles at different times; the arc with the largest radius corresponds to the earliest time and the smallest to the latest time. The particles that reach the $\Omega_{\alpha} \cap \Omega_{A}$ region will be removed from the energetic-$\alpha$ domain at a rate $\nu_{\alpha \rightarrow A}$ (to be discussed shortly).}
    \label{fig:dual_grid_sinking_of_energetic_particles}
\end{figure}

For the DT-$\alpha$ case, $\alpha$-particles in the intersection region with the ash grid will be removed from the energetic-$\alpha$ domain so as to prevent an under-resolved singular structure from forming without bound. Figure~\ref{fig:dual_grid_sinking_of_energetic_particles} illustrates this process.
The effect is modeled by introducing a sink term in the Fokker-Planck equation for the $\alpha$ species:
\begin{flalign}
    \label{eq:alpha_fp_wh_numerical_sink}
    &
    \frac{\partial f_{\alpha}}{\partial t} = \sum_{\gamma' \in {\cal G}} C_{\alpha \gamma'} +
    C^*_{\alpha  e} +
    S_{\alpha} - 
    S_{\alpha \rightarrow A}.
    &
\end{flalign}
Here, the sink term is defined as:
\begin{flalign}
    \label{eq:numerical_sink}
    &
    S_{\alpha \rightarrow A} = 
    \left\{
        \begin{array}{ccc}
            \nu_{\alpha \rightarrow A} \left({\bf v}\right) f_{\alpha},     &   \text{if}  &  {\bf v} \in \Omega_{A}, \\
            0,   & \text{otherwise},     &
        \end{array}
    \right.
    &
\end{flalign}
with $\Omega_{\alpha} = v_{\alpha,\parallel} \times v_{\alpha,\perp}  \subset \mathbb{R}^2$ and $\Omega_{A} = v_{A,\parallel} \times v_{A,\perp} \subset \mathbb{R}^2$ the finite velocity domains of $\alpha$ particles and ash species, respectively. The velocity-dependent sink rate, $\nu_{\alpha \rightarrow A} \left({\bf v}; \boldsymbol{\zeta}\right)$, is parameterized by $\boldsymbol{\zeta}$ (to be discussed shortly) and determined from the asymptotic behavior of collisions between energetic $\alpha$ particles and thermal ion species, $\gamma'$. Consider their collision operator:
\begin{flalign}
    \label{eq:alpha_thermal_ion_collision}
    &
    C_{\alpha \gamma'} = \Gamma_{\alpha \gamma'} \nabla_v \cdot \left[ {\cal D}_{\gamma '} \cdot \nabla_v f_{\alpha} - \frac{m_{\alpha}}{m_{\gamma'}} {\bf A}_{\gamma'} f_{\alpha} \right].
    &
\end{flalign}
At typical velocity scales of the energetic population, where \(v \sim \Delta v_{\alpha} \gg v_{th,\gamma'} \), we approximate the thermal distribution function as a Maxwellian, yielding:
\begin{flalign}
    \label{eq:alpha_thermal_ion_collision_approx_step0}
    &
    C_{\alpha \gamma'} \approx \Gamma_{\alpha \gamma'} \nabla_v \cdot 
    \left[ 
        {\cal D}^M_{\gamma'} \cdot \nabla_v f_{\alpha} - \frac{m_{\alpha}}{m_{\gamma'}} {\bf A}^M_{\gamma'} f_{\alpha}
    \right],
    &
\end{flalign}
where
\begin{flalign}
    \label{eq:alpha_thermal_ion_collision_approx_step1}
    &
    {\bf A}^M_{\gamma'}
    =
    \frac{n_{\gamma'}}{\pi^{3/2} v^3_{th,\gamma'}}
    \left[
        \frac{4 \pi v^2_{th,\gamma'} \text{exp}\left(-w^2/v^2_{th,\gamma'}\right)}{w}
        -
        \frac{2 \pi^{3/2} v^3_{th,\gamma'} \text{erf}\left( w / v_{th,\gamma'}\right)}{w^2}
    \right] 
    {\bf e}_{w},
    &
\end{flalign}
with $w^2 = \left|{\bf v} - {\bf u}_{\gamma'} \right|^2 = \left(v_{\|} - u_{\|,\gamma'} \right)^2 + v^2_{\perp}$, $v^2_{th,\gamma'} = 2 T_{\gamma'} / m_{\gamma'}$, and ${\bf e}_w = \frac{{\bf w}}{w}$. Further, due to the cold nature of the thermal species relative to $\alpha$ scale, the diffusion effect is negligible on the scale of $v \sim \Delta v_{\alpha} \gg v_{th,\gamma'}$ compared to collisional friction. Using the product rule, we get:
\begin{flalign}
    \label{eq:alpha_thermal_ion_collision_approx}
    &
    C_{\alpha \gamma'} 
    \approx 
    - \Gamma_{\alpha \gamma'} \frac{m_{\alpha}}{m_{\gamma'}} 
    \left[ 
        f_{\alpha} \nabla_v \cdot {\bf A}^{M}_{\gamma'}
        +
        {\bf A}^M_{\gamma'} \cdot \nabla_v f_{\alpha}
    \right].
    &
\end{flalign}
Recalling that \( {\bf A}^M_{\gamma'} = \nabla_v H^M_{\gamma'} \) we obtain:
\begin{flalign}
    \label{eq:expanded_approximated_collision_operator}
    &
    C_{\alpha\gamma'} \approx 
    -\Gamma_{\alpha\gamma'}\frac{m_{\alpha}}{m_{\gamma'}} 
    \left[
        f_{\alpha} \nabla^2_v H^M_{\gamma'}
        +
        {\bf A}^M_{\gamma'} \cdot \nabla_v f_{\alpha}
    \right].
    &
\end{flalign}
Using the Poisson equation for the $H^M_{\gamma'}$ potential, 
\begin{flalign}
    &
    \nabla^2_v H^M_{\gamma'} = -8 \pi f^M_{\gamma'},
    &
\end{flalign}
and multiplying and dividing the second term in the square bracket of Eq.~\eqref{eq:expanded_approximated_collision_operator} by \(f_{\alpha}\), we obtain:
\begin{flalign}
    \label{eq:expanded_approximated_collision_operator2}
    &
    C_{\alpha\gamma'} \approx
    \Gamma_{\alpha\gamma'} \frac{m_{\alpha}}{m_{\gamma'}}
    \left[
        8\pi f^M_{\gamma'}
        -
        {\bf A}^M_{\gamma'} \cdot \nabla_v \ln f_{\alpha}
    \right]f_{\alpha}.
    &
\end{flalign}
Finally, using the definition of a Maxwellian:
\begin{flalign}
    \label{eq:maxwellian_pdf}
    &
    f^M_{\gamma'}\left({\bf v} ; n_{\gamma'}, {\bf u}_{\gamma'}, T_{\gamma'}, m_{\gamma'} \right) 
    = 
    \frac{n_{\gamma'}}{\pi^{3/2}v^3_{th,\gamma'}} \exp\left( - \frac{w^2}{v^2_{th,\gamma'}} \right),
    &
\end{flalign} 
we obtain:
\begin{flalign}
    \label{eq:final_approximated_collision_operator}
    &
    C_{\alpha\gamma'} 
    \approx 
    \widetilde{\nu}_{\alpha\gamma'} 
    \left[
        \exp\left( - w^2/v^2_{th,\gamma'} \right)
        -
        \widehat{\bf A}^M_{\gamma'} \cdot \nabla_v \ln f_{\alpha}
    \right]
     f_{\alpha},
    &
\end{flalign}
where the effective collision frequency between the \(\alpha\) and $\gamma'$ species is:
\begin{flalign}
    &
    \widetilde{\nu}_{\alpha\gamma'} = \frac{Z^2_{\alpha}Z^2_{\gamma'}e^4 \ln\Lambda_{\alpha\gamma'}n_{\gamma'} m^{1/2}_{\gamma'}}{2 \sqrt{2} \epsilon^2_0 m_{\alpha} \pi^{3/2} T^{3/2}_{\gamma'}}.
    &
\end{flalign}
and
\begin{flalign}
    \label{eq:alpha_thermal_ion_collision_approx_Ahat}
    &
    \widehat{\bf A}^M_{\gamma'}
    =
    \left[
    \frac{v^2_{th,\gamma'} \text{exp}\left(-w^2/v^2_{th,\gamma'}\right)}{2w}
    -
    \frac{\sqrt{\pi} v^3_{th,\gamma'} \text{erf}\left( w / v_{th,\gamma'}\right)}{4 w^2}
    \right]
    {\bf e}_{w}.
    &
\end{flalign}
This reduction of the collision operator for the energetic-thermal-ion interaction thus resembles a {\em sink term}, which intuitively acts to remove energetic population from the overlapping region (preventing the singular structure from forming) and populate the thermal ash population. Note that the quantity could become a \textit{source} when $\widehat{\bf A}^M_{\gamma'} \cdot \nabla_v \ln{f_{\alpha}} > \exp\left(-w^2/v^2_{th,\gamma'}\right)$. In our work, we thus simplify the treatment by dropping the second term in Eq.~\eqref{eq:final_approximated_collision_operator} and therefore define the velocity-dependent sink rate for the energetic species in the overlapping region as:
\begin{flalign}
    \label{eq:vel_dep_sink_rate_energetic}
    &
    \nu_{\alpha\rightarrow A} \left( {\bf v} ; \boldsymbol{\zeta}\right) = \sum_{\gamma' \in {\cal G}^{\backslash \alpha}}  
    \widetilde{\nu}_{\alpha \gamma'} 
    \left[
        \exp\left( - w^2/v^2_{th,\gamma'} \right)
    \right],
    &
\end{flalign}
where ${\cal G}^{\backslash \alpha}$ is a shorthand notation for the ion species set excluding the $\alpha$ species, and the parameter vector is $\boldsymbol{\zeta} = \left\{ n_1, \cdots , n_{N_i}, {\bf u}_1,\cdots,{\bf u}_{N_i},T_1,\cdots, T_{N_i} \right\}$, where \( N_i = \text{length}\left({\cal G}\right) \) the total number of ion species.
The consistent handling of the second term will be left for future work. Dropping this term therefore affects only how the $^4$He population is partitioned between the energetic \(\alpha\) and thermal ash components, and not the global conservation of the combined population: because the sink of Eq.~\eqref{eq:vel_dep_sink_rate_energetic} is applied symmetrically as a source to the ash [Eq.~\eqref{eq:ash_fp_wh_numerical_source}], the total $^4$He moments are conserved exactly regardless of which terms are retained in the reduced operator. We emphasize, however, that because the sink is applied symmetrically, \( S_{A \leftarrow \alpha} = S_{\alpha \rightarrow A} \), the total $^4$He mass, momentum, and energy of the combined \(\alpha\)/ash population, \( f_{He} \), are conserved exactly for any choice of sink model.

The energetic particle population lost in the intersection region due to the sink term is subsequently transferred to the thermal ash population by modifying the Fokker-Planck equation for the ash:
\begin{flalign}
    \label{eq:ash_fp_wh_numerical_source}
    &
    \frac{\partial f_{A}}{\partial t} = \sum_{\gamma' \in {\cal G}} C_{A \gamma'} + C_{Ae} + S_{A \leftarrow \alpha},
    &
\end{flalign}
where \( S_{A \leftarrow \alpha} ({\bf v}) = S_{\alpha \rightarrow A} ({\bf v}) \) represents the numerical source for the ash population, originating from the numerical sink of the \(\alpha\) species. In the continuum limit, they satisfy a detailed balance relationship, ensuring consistency with the original PDE. This can be readily verified by defining the total $^4$He distribution function as the sum of the energetic \(\alpha\) and ash species, \( f_{He} := f_{\alpha} + f_{A} \), and adding Eqs.~\eqref{eq:ash_fp_wh_numerical_source} and \eqref{eq:alpha_fp_wh_numerical_sink}:
\begin{flalign}
    \label{eq:proof_of_detailed_balance}
    &
    \frac{\partial f_{He}}{\partial t} = \sum_{\gamma' \in {\cal G}} C_{He\gamma'} + C^*_{He,e} + S_{\alpha},
    &
\end{flalign}
which recovers the original single-population Fokker-Planck equation for the combined $^4$He (\(\alpha\)+ash) population. Consequently, the new formulation remains valid for arbitrary values of \( v^*_{\alpha} / v_{th,A} \), in contrast to Ref.~\onlinecite{peigney2014alpha}, which relies on asymptotic scale separation arguments and is strictly applicable only when \( v^*_{\alpha} / v_{th,A} \gg 1 \). This distinction is crucial, particularly in scenarios involving strong spatial heterogeneity due to shock waves, but also where \(\alpha\)-particle energy deposition can dynamically heat the background plasma---potentially violating the assumption. In radiation-hydrodynamic simulations of NIF ignition experiments\cite{haines_2024_pop_ignition_simulation}, fuel temperatures exceeded 30 keV, leading to \( v_{th,A} \sim 0.5 v^*_{\alpha} \), barely satisfying the required scale separation. Notably, Eulerian approaches for modeling ICF plasmas often necessitate velocity space grids spanning several thermal speeds, rendering the fuel and ash computational domains comparable to or larger than the \(\alpha\)-particle grid---thereby challenging the scale separation assumption at the distribution tail.

Although the numerical sink term, $S_{A\leftarrow\alpha}$, prevents the runaway accumulation of particles near \( v \sim v_{th,A} \), unresolved features may still emerge, particularly in the presence of cold materials (e.g., DT ice). To regularize the solution, we follow a similar strategy to Ref.~\onlinecite{daniel_ccp_2020_relativistic_fp_0d2v} which mollified the mathematical singularity arising in the Lorentz collision operator as $v \rightarrow 0$. We modify the energetic collision operator with fluid electrons:
\begin{flalign}
    \label{eqn:ie_fp_modified}
    &
    C^*_{\alpha e} = \nu_{\alpha e} \nabla_v \cdot \left[ 
    \frac{T_e + \delta T_{\alpha}}{m_{\alpha}} \nabla_v f_{\alpha} - \left({\bf u}_{\alpha} - {\bf v} \right)f_{\alpha}
    \right].
    &
\end{flalign}
Here, \( \delta T_{\alpha} \in \mathbb{R}_{+} \) serves as a {\it floor} temperature for the collisional diffusion coefficient with fluid electrons. 
In this study, we define it based on the minimum grid resolution, ensuring that in the continuum limit, \( \delta T_{\alpha} \rightarrow 0 \), thus recovering consistency with the original equation. This floor temperature effectively dissipates singular structures in the distribution function near \( v \sim v_{th,A} \) over a few computational cells.

To ensure that this regularization term does not introduce additional source of energy, we also modify the fluid electron equation:
\begin{flalign}
    \label{eqn:fluid_e_modified}
    &
    \frac{3}{2}\frac{\partial }{\partial t}\left(n_e T_e\right) = 
    3\sum_{\gamma \in {\cal G}^{\backslash \alpha}} n_e \nu_{e\gamma} \left(T_{\gamma} - T_{e} \right) +
    3n_e \nu_{e\alpha} \left( T_{\alpha} - T_e - \delta T_{\alpha}\right).
    &
\end{flalign}
The detail of how this preserves continuum moments of the distribution function is left for Appendix~\ref{app:moment_cons_thybridfloor}.
In this study, we discretely solve Eqs.~\eqref{eq:ash_fp_wh_numerical_source} for the ash species, \eqref{eq:alpha_fp_wh_numerical_sink} for the energetic \(\alpha\) species, \eqref{eqn:vfp} for all other ions (with $S_{\gamma} = 0$ for them), and \eqref{eqn:fluid_e_modified} for fluid electrons. The discrete implementation utilizes the {\it discrete nonlinear constraint strategy}, as detailed in Ref.~\onlinecite{taitano_jcp_2018_vfp}, to enforce conservation. The discrete fluid electron equation and energetic-fluid-electron collision operator are modified accordingly. We discuss next the discrete treatment of these equations and their various terms.
%
%
%
%
%
%
\section{Numerical Implementation}
\label{sec:numerical_implementation}
As in earlier studies\cite{taitano_cpc_2021_ifp_code, taitano_jcp_2015_rfp}, we consider a conservative finite difference formulation. First, we define the parallel velocity grid as
\(\widehat{W}_{\|} = \left\{-\widehat{w}_{\|,\max} + \left(j - \frac{1}{2}\right) \Delta \widehat{w}_{\|} \right\}_{j=1}^{N_{\|}}\),
and the perpendicular velocity grid as \(\widehat{W}_{\perp} = \left\{\left(k - \frac{1}{2}\right) \Delta \widehat{w}_{\perp} \right\}_{k=1}^{N_{\perp}}\).
Here, the grid spacing in the transformed parallel velocity coordinate is
\(\Delta \widehat{w}_{\|} = \frac{2\widehat{w}_{\|,\max}}{N_{\|}}\),
and the grid spacing in the transformed perpendicular velocity coordinate is
\(\Delta \widehat{w}_{\perp} = \frac{\widehat{w}_{\perp,\max}}{N_{\perp}}\),
where \(\widehat{w}_{\|,\max}\) and \(\widehat{w}_{\perp,\max}\) denote the maximum limits of the parallel and perpendicular velocity domains, respectively, and \(N_{\|}\) and \(N_{\perp}\) denote the corresponding numbers of grid points.

The transformed cylindrical velocity-space domain is then constructed as the tensor product
\(
\Omega_{\widehat{w}} = \widehat{W}_{\|} \times \widehat{W}_{\perp}.
\)
For a given species \(\gamma\), the mapping to the original coordinate system is defined by
\(
V_{\|,\gamma} = v^*_{\gamma} \, \widehat{W}_{\|} + {\bf e}_{\|}\, u^*_{\|,\gamma}, \quad \text{and} \quad V_{\perp,\gamma} = v^*_{\gamma} \, \widehat{W}_{\perp},
\)
with the associated domain
\(
\Omega_{v,\gamma} = V_{\|,\gamma} \times V_{\perp,\gamma}.
\)
The distribution function for species \(\gamma\) on the cell \(\widehat{\bf w}_{j,k} = \left\{ \widehat{w}_{\|,j}, \widehat{w}_{\perp,k} \right\}\) is then defined as
\(
f_{\gamma,j,k} \equiv f_{\gamma}\left( \widehat{w}_{\|,j}, \widehat{w}_{\perp,k}\right).
\)
Discrete velocity-space moments over the species \(\gamma\) domain, between two functions \(\psi({\bf v})\) and \(\phi({\bf v})\), are computed using a midpoint quadrature rule as follows:
\begin{flalign}
    \label{eq:moments_definition}
    &
    \left< \psi, \phi \right>_{\delta v_{\gamma}} = 2\pi v_{\gamma}^{*3} \sum_{j=1}^{N_{\|}} \Delta \widehat{w}_{\|} \sum_{k=1}^{N_{\perp}} \widehat{w}_{\perp,k} \Delta \widehat{w}_{\perp} \, \psi_{j,k} \, \phi_{j,k}.
    &
\end{flalign}
Further, unless otherwise mentioned, all distribution functions are initialized as Maxwellians, parameterized in terms of the prescribed number density, drift velocity, and temperature,
\begin{flalign}
    \label{eq:maxwellian_normalized}
    &
    f_M \left( \widehat{w}_{\|}, \widehat{w}_{\perp} ; n_0, u_{\|}, v_{th} \right) = 
    \frac{n_0}{\pi^{3/2} v^3_{th}} 
    \exp\left[{-\frac{\left(\widehat{w}_{\|}v^* + u^*_{\|} - u_{\|} \right)^2 + \left(\widehat{w}_{\perp}v^*\right)^2}{v^2_{th}}}\right],
    &
\end{flalign}
where we have dropped the explicit species index for brevity.

We employ a nonlinearly implicit time-stepping strategy that iterates the solution to convergence at each time-step. This is motivated by the fact that, despite the dynamical time-scale induced by the fusion reaction rate, 
\(
\nu_{DT\rightarrow\alpha} \sim \left|\frac{S_{\alpha}}{f^{l+1}_{\alpha} - f^{l}_{\alpha}}\right|
\)
(where the superscript \(l\) denotes the time-step) not being stiff, the collisions and {\em numerical sink} terms are, due to the scaling \(\Delta t \, \nu_{\alpha e} \gg 1\) in realistic problems involving dense ice layers and shells encountered in ICF experiments. Furthermore, we require that a wide variety of critical physical symmetries and a tight balance between terms be maintained, including conservation symmetries and equilibration toward the asymptotic Maxwellian equilibrium, which are often nonlinear with respect to the solution variable.

In this work, we use a second order backward differencing time integrator (BDF2). Accordingly, the fully discretized equations for the energetic population distribution function are given as:
\begin{flalign}
    &
    \sum^{1}_{l'=-1}c^{l+l'} f^{l+l'}_{\alpha,j,k}
    - 
    \Delta t^{l} 
    \left\{ \sum_{\gamma' \in {\cal G}} C^{l+1}_{\alpha \gamma',j,k}
    +
    C^{*^{l+1}}_{\alpha  e, j,k} 
    +
    S^{l+1}_{\alpha, j,k} 
    - 
    S^{l+1}_{\alpha \rightarrow A, j,k} 
    \right\} = 0,
    &
    \label{eq:discrete_alpha_fp_wh_numerical_sink}
\end{flalign}
the ash population distribution function as:
\begin{flalign}
    \label{eq:discrete_ash_fp_wh_numerical_source}
    &
    \sum^{1}_{l'=-1}c^{l+l'} f^{l+l'}_{A,j,k}
    - 
    \Delta t^{l} \left\{  
    \sum_{\gamma' \in {\cal G}} C^{l+1}_{A\gamma',j,k} 
    + 
    C^{l+1}_{Ae,j,k} 
    + 
    S^{l+1}_{A \leftarrow \alpha,j,k}
    \right\} = 0,
    &
\end{flalign}
the rest of ion distribution functions, $\gamma \notin \left\{ \alpha, A \right\} $, as:
\begin{flalign}
    \label{eq:discrete_ion_fp}
    &
    \sum^{1}_{l'=-1}c^{l+l'} f^{l+l'}_{\gamma,j,k}
    - 
    \Delta t^{l} \left\{  
    \sum_{\gamma' \in {\cal G}} C^{l+1}_{\gamma \gamma',j,k} 
    + 
    C^{l+1}_{\gamma e,j,k}
    \right\} = 0,
    &
\end{flalign}
and the fluid electron equation as:
\begin{flalign}
    &
    \frac{3}{2}\sum^{1}_{l'=-1}c^{l+l'} n^{l+l'}_e T^{l+l'}_e
    -
    3\Delta t^{l}
    \sum_{\gamma \in {\cal G}} n^{l+1}_e \nu^{l+1}_{e\gamma} \left(T^{l+1}_{\gamma} - T^{l+1}_{e} - \delta T_{\gamma} \right)
    = 0.
    \label{eqn:discrete_fluid_e_modified}
    &
\end{flalign}
Here, the floor temperature is applied only to the \(\alpha\) species, i.e., \( \delta T_{\gamma} := \delta T_{\alpha}\delta_{\gamma\alpha} \) with \(\delta_{\gamma\alpha}\) the Kronecker delta (see Appendix~\ref{app:chang_cooper}), so that it vanishes for all non-\(\alpha\) species. Here, \( c^l \) denotes the backward differencing coefficient at the \( l^{\text{th}} \) time level\cite{byrne_bdf2_1975}, and the solution vector is defined as
\(
\mathbf{x}^l = \{ \mathbf{f}^l, T_e^l \} \in \mathbb{R}_+^{N_x},
\)
where \(\mathbf{f} = \{ \mathbf{f}_1, \cdots, \mathbf{f}_{N_i} \} \in \mathbb{R}_+^{N_i N_v}\) and
\(
\mathbf{f}_\gamma = \{ f_{\gamma,1,1}, \cdots, f_{\gamma,N_{\|},1}, f_{\gamma,1,2}, \cdots, f_{\gamma,N_{\|},N_{\perp}} \} \in \mathbb{R}_+^{N_v}
\)
represents the \(\gamma\) ion distribution function. Here, \( T_e \in \mathbb{R}_+ \) is the fluid electron temperature, \( N_v = N_{\|}N_{\perp} \) is the total number of grid points in velocity space, and \( N_x = N_i N_v + 1 \) is the total number of unknowns in the system.

A detailed discussion of the discrete treatment of the ion-ion collision operator is beyond the scope of this work; interested readers are referred to Refs.~\onlinecite{taitano_jcp_2015_rfp, taitano_cpc_2021_vth_u_shift, taitano_jcp_2016_vgrid_adaptivity, taitano_jcp_2017_ep_rfp, anderson_jcp_2020_adaptive_grid} for further details. Moreover, the stiff collision terms are addressed using a specialized variant of a high-order/low-order (HOLO) algorithm\cite{chacon_jcp_2017_holo_review} adapted for the RFP equations. Although this strategy is critical for overcoming the fast collision time of the DT ice-layer---a scenario that will be explored in subsequent numerical examples---the details are deferred to future publications. Instead, the following subsections focus on the discretization of \( C^{*}_{\alpha e,j,k} \), \( S_{\alpha,j,k} \), \( S_{\alpha\rightarrow A,j,k} \), and \( S_{A\leftarrow\alpha,j,k} \). Unless stated otherwise, the explicit time-indexing is omitted from this point on for brevity.
%
%
%
\subsection{Treatment of the source for the energetic population, $S_{\alpha,j,k}$}
\label{sec:treament_of_energetic_source}
The source of energetic particles, particularly those generated by exothermic reactions such as $D + T \rightarrow \alpha \;(3.5\,\text{MeV}) + n \;(14.1\,\text{MeV})$, often exhibits thermal broadening in its energy spectrum due to the finite temperature of the reactant populations. However, this broadening is extremely difficult to resolve with finite resolution (e.g., on the order of $\sim$keV within an $\sim$MeV energy grid) and thus appears numerically as an effective delta function. To mitigate grid-aliasing effects caused by sharp velocity-space features introduced by the source, its definition will be regularized using a Gaussian broadening centered around the mean product speed, $v_{\alpha}$, with a finite variance proportional to the grid size: 
\begin{flalign}
    \label{eq:discrete_source_definition}
    &
    S_{\alpha,j,k} = \frac{\widetilde{R}_{DT}}{4\pi v^2_{j,k} \sqrt{\pi \Delta v^2_{\alpha}}} \exp{\left[ - \frac{\left(v_{j,k} - v_{\alpha} \right)^2}{\Delta v^2_{\alpha}} \right]},
    &
\end{flalign}
where, 
\begin{flalign}
    \label{eq:grid_magnitude}
    &
    \Delta v_{\alpha} = v^*_{\alpha}\Delta\widehat{w} = v^*_{\alpha}\sqrt{\Delta \widehat{w}_{\|}^2 + \Delta \widehat{w}_{\perp}^2} = \sqrt{\Delta v_{\alpha,\|}^2 + \Delta v_{\alpha,\perp}^2}
    &
\end{flalign}
and $v_{j,k} = v^*_{\alpha}\sqrt{\widehat{w}^2_{\|,j} + \widehat{w}^2_{\perp,k}}$. The component grid spacings appearing in Eq.~\eqref{eq:grid_magnitude} are $\Delta v_{\alpha,\|} = v^*_{\alpha}\Delta\widehat{w}_{\|}$ and $\Delta v_{\alpha,\perp} = v^*_{\alpha}\Delta\widehat{w}_{\perp}$. Here,
\begin{flalign}
    \label{eq:source_scaling}
    &
    \widetilde{R}_{DT} = \eta R_{DT}
    &
\end{flalign}
and 
\begin{flalign}
    \label{eq:scaling_factor}
    &
    \eta = \left\{
        \sum^{N_{\|}}_{j} \sum^{N_{\perp}}_{k} \frac{\Delta v_{\alpha,\|} \Delta v_{\alpha,\perp} v_{\alpha,\perp,k}}{2 v^2_{j,k} \sqrt{\pi \Delta v^2_{\alpha}}} \exp{\left[ - \frac{\left(v_{j,k} - v_{\alpha} \right)^2}{\Delta v^2_{\alpha}} \right]}
    \right\}^{-1}
    &
\end{flalign}
is the numerical normalization constant that ensures $\left<1,{\bf S}_{\alpha} \right>_{\delta v_{\alpha}} = R_{DT}$, where ${\bf S}_{\alpha} = \left\{S_{\alpha,1,1} , \cdots, S_{\alpha,N_{\|},N_{\perp}} \right\}$. It can be trivially seen that the source recovers the Dirac delta function form in the continuum as $\Delta v \rightarrow 0$. It should be noted that this non-smooth behavior of the source can impact the convergence rate of the solution as the velocity grid is refined. When this is a concern, the source width dependent on grid spacing can be added in quadrature with a fixed source width independent of the grid refinement to ensure a smooth source. 
\subsection{Treatment of the energetic sink to ash, $S_{\alpha\rightarrow A}$}
\label{sec:treament_of_sink}

Once reaction product particles in the energetic/$\alpha$ grid slow down to thermal energies, they are transferred to the thermal population via a sink term:

\begin{flalign}
    \label{eq:discrete_energetic_to_ash_sinking}
    &
    S_{\alpha\rightarrow A,j,k} = 
    \beta
    \left\{
        \begin{array}{ccc}
            \left< \nu_{\alpha\rightarrow A} \right>_{j,k} f_{\alpha,j,k} & \text{if} &  \Omega_{\alpha,j,k} \cap \Omega_A \neq \emptyset,  \\
            0 & \text{otherwise} & 
        \end{array}
    \right\}.
    &
\end{flalign}
Here, \(\Omega_{\alpha,j,k} = \left[v_{\alpha,\|,j-\frac{1}{2}}, v_{\alpha,\|,j+\frac{1}{2}} \right] \times \left[ v_{\alpha,\perp,k-\frac{1}{2}}, v_{\alpha,\perp,k+\frac{1}{2}} \right]\) represents the domain of cell \(\left(j,k\right)\), and \(\emptyset\) denotes the null set. The prefactor \(\beta\)
\begin{flalign}
    &
    \beta = \frac{1}{2} \left( \tanh \left[ \chi \left( \log_{10} \frac{n_{\alpha}}{n_{min}} \right) \right] + 1 \right )
    &
\end{flalign}
serves as an {\it activation function} to prevent numerical issues arising when the energetic population is driven to zero. Such challenges occur when the plasma remains cold, efficiently slowing down and draining the energetic \(\alpha\) particles, but not supporting fusion reactions locally (e.g., DT ice regions). Here, \(\chi\) is the hyperbolic smoothing factor (empirically set to 2 in this study unless otherwise specified), \(n_{\alpha} = \left< 1, f_{\alpha} \right>_{\delta v}\) denotes the energetic number density, and \(n_{min}\) represents the threshold number density for sink-term cutoff. A smooth activation function is necessary to maintain differentiability with respect to the solution variables, ensuring robust convergence of our nonlinearly implicit iterative solver. 

The averaged sinking rate in Eq.~\eqref{eq:discrete_energetic_to_ash_sinking} is given by:
\begin{flalign}
    \label{eq:averaged_sinking_rate}
    &
    \left< \nu_{\alpha\rightarrow A} \right>_{j,k} = 
    \sum_{\gamma' \in {\cal G}^{\backslash \alpha}} \widetilde{\nu}_{\alpha\gamma'} 
        \left< \exp\left( - \frac{w^2_{\gamma'}}{v^2_{th,\gamma'}} \right) \right>_{j,k}
    ,
    &
\end{flalign}
where the local velocity-space averaging operator, 
\(\left< \circ \right>_{j,k} = \frac{\int_{\Omega_{\alpha,j,k}} dv_{\perp} v_{\perp} dv_{\|} \circ}{\int_{\Omega_{\alpha,j,k}} dv_{\perp} v_{\perp} dv_{\|}} \) 
yields the following explicit formula:
\begin{flalign}
    \label{eq:averaging_definition}
    &
    \left< \exp\left( - \frac{w^2_{\gamma'}}{v^2_{th,\gamma'}} \right) \right>_{j,k} =
    -\frac{
        \sqrt{\pi} v^3_{th,\gamma'}
        \left[
        \text{erf} \left(\frac{w_{\|,\gamma',j+\frac{1}{2}}}{v_{th,\gamma'}}\right)
        -
        \text{erf} \left(\frac{w_{\|,\gamma',j-\frac{1}{2}}}{v_{th,\gamma'}}\right)
        \right]
        \left[
            \exp \left(-\frac{v^2_{\perp,k+\frac{1}{2}}}{v^2_{th,\gamma'}} \right)
            -
            \exp \left(-\frac{v^2_{\perp,k-\frac{1}{2}}}{v^2_{th,\gamma'}} \right)
        \right]
    }
    {2\left(v_{\|,j+\frac{1}{2}} - v_{\|,j-\frac{1}{2}} \right)\left(v^2_{\perp,k+\frac{1}{2}} - v^2_{\perp,k-\frac{1}{2}} \right)}.
    &
\end{flalign}
The expression represents the analytical coarse-graining in cell \(\left(j,k\right)\). Coarse-graining is necessary because the thermal/ash population is often under-resolved on the energetic/$\alpha$ grid.
%
%
%
\subsection{Treatment of the ash source from energetic species, $S_{A \leftarrow \alpha}$}
\label{sec:treament_of_source}
The sink from the energetic species grid, $S_{\alpha \rightarrow A}$, must be reconstructed onto the ash-species grid. To ensure that this reconstruction maintains the monotonicity of the original source, we employ a bilinear interpolation strategy:
\begin{flalign}
    \label{eq:sink_reconstruction_bli}
    &
    S_{A \leftarrow \alpha,j,k} = {\cal L}_{j,k} {\bf S}_{\alpha \rightarrow A},
    &
\end{flalign}
where $S_{A \leftarrow \alpha,j,k}$ represents the source reconstruction on the $\left(j,k\right)$ velocity space grid for the ash species. The function ${\cal L}_{j,k} : \mathbb{R}^{N_v}_+ \rightarrow \mathbb{R}_+$ is the bilinear interpolation map at the $\left(j,k\right)$ ash grid point, and ${\bf S}_{\alpha\rightarrow A} = \left\{ S_{\alpha\rightarrow A,1}, \dots, S_{\alpha \rightarrow A,N_v} \right\} \in {\mathbb R}^{N_v}_{+}$ is the vectorized representation of the original source on $\Omega_{v,\alpha}$. 

The interpolation operation does not strictly preserve the discrete velocity-space moments of the sink; rather, it ensures preservation only up to the order of the interpolation and quadrature scheme used, i.e.,
\begin{flalign*}
    &
    \left< \phi , S_{A\leftarrow\alpha} \right>_{\delta v_{A}} = \left< \phi , S_{\alpha \rightarrow A} \right>_{\delta v_{\alpha}} + {\cal O} \left( \Delta \widehat{w} ^2 \right).
    &
\end{flalign*}
Here, $\phi \left(v^*_{\alpha} \widehat{\bf w} + {\bf u}^*_{\alpha}\right)$ is the velocity-space function used in the inner product evaluation. In our approach, the total mass, momentum, and energy of the sink, $S_{\alpha\rightarrow A}$, corresponding to $\phi \in \left\{1, v_{\|}, \frac{v^2}{2} \right\}$, are exactly balanced through the reconstructed source, $S_{A\leftarrow\alpha}$, by introducing an additive correction to project out the errors:
\begin{flalign}
    \label{eq:sink_reconstruction_plus_correction}
    &
    S_{A\leftarrow\alpha,j,k} := S_{A\leftarrow\alpha,j,k} (1 + \sum^2_{l=0} C_l \psi_{l,j,k})
    &
\end{flalign}

Here, the coefficient $C_l$ is the $l^{th}$ projection coefficient, describing the coupling to the projection function $\psi_{l,j,k} = \psi_l\left(\widehat{\bf w}_{j,k} \right)$. This term ensures the discretization preserves certain moments when moving particles from the alpha to ash grid; details are given in Appendix~\ref{app:moment_cons_source}.
%
%
%

\subsection{Treatment of the Energetic-Species-Fluid-Electron Collision term $C^*_{\alpha e}$ and the Fluid Electron Equation}
\label{sec:treament_of_fluid_electrons}
For ion/fluid-electron interactions, modeled using a Lenard-Bernstein form, we adopt the Chang-Cooper discretization of the velocity-space flux\cite{chang_jcp_1970_chang_cooper}, which is briefly summarized in Appendix~\ref{app:chang_cooper}. This discretization is designed to preserve Maxwellian distributions and will tend to drive the distribution function towards such a state. 

For energetic particles, however, this Maxwellian and its associated thermal speed often fall below the grid scale, i.e., $v_{th} = \sqrt{2T_e/m_{\alpha}} \ll \Delta v_{\alpha}$, where $\Delta v_{\alpha} = v^*_{\alpha} \Delta \widehat{w}$. This may result in a singular structure that poses difficulties for naive numerical treatments of $\alpha$-electron interactions, although these challenges are significantly mitigated by the numerical sink formulation described in Section~\ref{sec:treament_of_sink}. To further mitigate this issue, we set the floor temperature introduced in Eq.~\eqref{eqn:ie_fp_modified} to a value related to the alpha velocity grid size:
\begin{flalign}
    \label{eq:delta_T_floor_def}
    &
    \delta T_{\alpha} = m_{\alpha} \Delta v_{\alpha}^2.
    &
\end{flalign}
This floor temperature ensures that the distribution function is smoothed across a single cell near $v\sim 0$. 

The potential impact of $\delta T_{\alpha}$ away from $v \approx 0$ can be assessed by considering the grid-scale diffusion and advection time scales associated with $C^*_{\alpha e}$ (assuming ${\bf u}_{\alpha} = {\bf 0}$ for simplicity):
\begin{flalign}
    \label{eq:diff_time_scale}
    &
    \tau_{diff} = \frac{m_{\alpha} \Delta v^2_{\alpha}}{ T_e + \delta T_{\alpha}  } \approx \frac{m_{\alpha} \Delta v^2_{\alpha}}{\delta T_{\alpha}  } = 1,
    &
\end{flalign}
and
\begin{flalign}
    \label{eq:adv_time_scale}
    &
    \tau_{adv,k} = \frac{\Delta v_{\alpha}}{v_{k}} = \frac{\Delta v_{\alpha}}{k \Delta v_{\alpha}} = \frac{1}{k}.
    &
\end{flalign}
Here, $k \in \left\{1,\dots,N_{v}\right\}$ is the grid index along the velocity magnitude coordinate, with $v_{max} = \Delta v_{\alpha} N_v$ and $N_v$ denoting the number of velocity grid points introduced for analysis purposes. Consequently, the relative strength of advection versus diffusion effects can be expressed as the ratio of diffusion to advection time scales:
\begin{flalign}
    \label{eq:ratio_lb_adv_2_diff}
    &
    \frac{\tau_{diff}}{\tau_{adv,k}} \approx k.
    &
\end{flalign}
Thus, the influence of diffusion from $\delta T_{\alpha}$ diminishes as we move away from $v \approx 0$ (or $k > 1$). Moreover, since grid-scale structures represent the finest resolution supported by the numerical system, the relative impact of flooring discussed here represents the worst-case scenario. In practical computations, non-physical effects are rarely observed, while the regularization of the inherent numerical singularity near the ash overlap region provides significant benefits. This will be demonstrated in subsequent sections, where numerical tests confirm that the choice of $\delta T_{\alpha}$ in Eq.~\eqref{eq:delta_T_floor_def} is sufficient to smooth singular structures near the origin.

%
%
\section{Numerical Results}
\label{sec:numerical_results}
The proposed multiscale algorithm is tested on a series of benchmark problems with increasing complexity. We first study the isolated 0D2V (zero spatial and two velocity-space dimensions; likewise 1D2V denotes one spatial and two velocity-space dimensions) system and compare the asymptotic tail structure and heating rate of electrons due to the fixed source of $\alpha$ particles against semi-analytical reference solutions to verify and benchmark our multiscale algorithm. We proceed by investigating the slowing down process in a 1D2V spatially heterogeneous setting to benchmark our implementation against a semi-analytical characteristics-based reference solution to verify the machinery in spherical geometry, followed by a convergence study in configuration and velocity space, and time to demonstrate the consistency of our approach with the continuum system. Finally, we demonstrate the proposed algorithm with a spherically imploding capsule with $\alpha$ production that mimics a layered ICF experiment. The algorithm has been implemented into the iFP coupled multiphysics VFP spherical implosion code\cite{taitano_cpc_2021_ifp_code, taitano_cpc_2021_vth_u_shift, taitano_jcp_2015_rfp, taitano_jcp_2016_vgrid_adaptivity, taitano_jcp_2017_ep_rfp, anderson_jcp_2020_adaptive_grid, hammer_2019_tans_ifp_and_radiation} which has been thoroughly verified and applied to both basic high energy density physics research\cite{keenan_pre_2017_deciphering_shock_structure, keenan_pop_2018_species_stratification, keenan_pop_2020_revolver_sim, keenan_pop_2020_shock_driven_kinetic_mix, andeste_pre_2021_fully_kinetic_shock, chu_prl_2023_colliding_shocks} as well as modeling ICF experiments\cite{taitano2018popomega, mannion_pre_2023_above_balabio, gatujohnson2024pre, reichelt2025pre_mix}. For all simulations, given the proton mass as $m_p = 1.6726\times10^{-27}$kg and the elementary charge as $e = 1.6022\times10^{-19}$C, the masses of ash/$\alpha$, electrons, deuterons, and tritons were set as $m_{A} = m_{\alpha} = 4m_p$, $m_e = m_p/1836$, $m_D = 2m_p$, and $m_T = 3m_p$ while the charges were set to $q_A = q_{\alpha} = 2e$, $q_e = -e$, $q_D = q_T = e$, respectively. 
%
%
%
%
%
%
\subsection{0D2V $\alpha$ Particle Slowing Down}
\label{subsec:0d2v_slowing_down}
\begin{figure}[t!]
    \centering
    \includegraphics[scale=0.7]{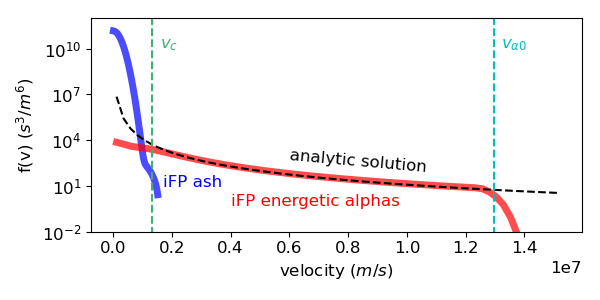}
    \caption{0D2V slowing-down distribution: The quasi-steady-state distribution function of ash (blue), \(\alpha\) particles (red), and the semi-analytical reference solution (black dashed line). The values of \( v_c \) and \( v_{\alpha} \) are indicated by the green and turquoise dashed lines, respectively.}
    \label{fig:0d2v_slowing_down}
\end{figure}

The 3.5 MeV \(\alpha\) particles (\(^4\mathrm{He}\)) produced in the deuterium-tritium (DT) fusion reaction are born at a velocity \( v_{\alpha} \). Their birth speed lies between the thermal speeds of the fuel ions and electrons, i.e., 

\begin{flalign*}
    &
    v_{th,i} = \sqrt{\frac{2T_i}{m_i}} \ll v_{\alpha} \lessapprox v_{th,e} = \sqrt{\frac{2T_e}{m_e}}.
    &
\end{flalign*}
This allows us to derive a simplified form of the collision operator and, consequently, a semi-analytical model for comparison with our multiscale two-grid scheme. We consider a spatially homogeneous scenario with a normalized velocity space domain of \( \left( \widehat{w}_{\|}, \widehat{w}_{\perp} \right) \in [-7, 7] \times \left[ 0, 7 \right] \), using fixed scaling and shift velocities given by 
\(
\left( v^*_{A}, v^*_{\alpha} \right) = \left( \qty{219}{\um\per\ns}, \qty{2190}{\um\per\ns} \right)
\)
and 
\(
\left( u^*_{\|,A}, u^*_{\|,\alpha} \right) = (0,0).
\)
The source strength is set to \( S_{0} = \qty{1.52e18}{cm^{-3} / ns} \), with a background ash density of \( n_{A,0} = \qty{1e22}{cm^{-3}} \) and a temperature of \( T_{A,0} = T_{e,0} = \qty{1}{\keV} \). The semi-analytical steady-state \(\alpha\)-particle distribution function is given by\cite{gaffey_energetic_ion_dist_1976}:
\begin{flalign}
    \label{eq:0d2v_semi_analytical_solution}
    &
    f^{\infty}_{\alpha} ( v ) = 
    \frac{S_0 \tau_{s} }{4\pi \left( v^3 + v^3_c \right)}, 
    \quad v \in \left[ v_c , v_{\alpha}  \right],
    &
\end{flalign}
where \( S_{0} \) is the fixed source of \(\alpha\) particles (equivalent to $R_{DT}$) and \( v \) is the particle speed. The critical speed, at which the collisional drag (energy loss) on the thermal ion species equals that on the electrons, is defined as:
\begin{flalign}
    \label{eq:critical_velocity}
    &
    v^3_c = \frac{3\sqrt{\pi} T^{3/2}_{e}}{\sqrt{2 m_e} n_e} \sum_{\gamma \in {\cal G}^{\backslash \alpha}} \frac{Z^2_{\gamma} n_{\gamma}}{m_{\gamma}}.
    &
\end{flalign}
The \(\alpha\) slowing-down time is given by
\begin{flalign}
    \label{eq:tau_s}
    &
    \tau_{s} = \nu_{\alpha e}^{-1} = \frac{6\sqrt{2}\,\pi^{3/2}\epsilon_0^2\, m_{\alpha} T_e^{3/2}}{Z_{\alpha}^2 n_e e^4 \ln\Lambda_{\alpha e}\sqrt{m_e}},
    &
\end{flalign}
where \( m_{\alpha} \) and \( Z_{\alpha} \) are the mass and charge number of the \(\alpha\) particles, \( m_e \) and \( n_e \) are the electron mass and number density, and \( \nu_{\alpha e} \) is the \(\alpha\)-electron collision frequency of Eq.~\eqref{eqn:nu_ie}. 
Figure~\ref{fig:0d2v_slowing_down} compares the \(\alpha\) particle distribution function obtained from our scheme with the semi-analytical asymptotic model.
As seen in Figure~\ref{fig:0d2v_slowing_down}, the \(\alpha\) distribution function correctly follows the expected \( 1/(v^3 + v^3_c) \) trend in the region \( v \in \left[ v_{c}, v_{\alpha} \right] \) and flattens below \( v_c \), where ion drag dominates the slowing down. The particles removed by the sink accumulate in the ash population, which remains Maxwellian at the background temperature; for the present parameters, \( v_c \approx 6\, v_{th,A} \), so the Maxwellian ash occupies only the region \( v \lesssim \) a few \( v_{th,A} \), well below \( v_c \). Since iFP evolves the distribution function on a 2V cylindrical domain, the results in Figure~\ref{fig:0d2v_slowing_down} were obtained by first mapping the distribution function to spherical coordinates and then averaging over the polar angle.

To evaluate the impact of our solution regularization strategy---where a numerical floor temperature, \( \delta T_{\alpha} \), is applied to the \(\alpha\)-electron collisional diffusion coefficient, as described in Sections~\ref{sec:conservative_multiscale_2_grid_strategy} and~\ref{sec:treament_of_fluid_electrons}---we compare the steady-state \(\alpha\)-distributions with and without temperature flooring. The simulations use the following initial conditions: 

\[
n_{e,0} = \qty{2.52e24}{cm^{-3}}, \quad 
n_{\alpha,0} = \qty{5.04e14}{cm^{-3}}, \quad 
n_{A,0} = \qty{1.26e24}{cm^{-3}}, \quad 
\]

\[
u_{\|,e} = 0, \quad 
u_{\|,\alpha} = 0, \quad 
u_{\|,A} = 0, \quad 
S_{0} = \qty{2.16e21}{cm^{-3}/ns}.
\]
The initial ash and electron temperatures are set to 50 eV to emulate conditions found as alphas stop in the cold dense fuel. 

\begin{figure}[h!]
    \centering
    \includegraphics[scale=0.6]{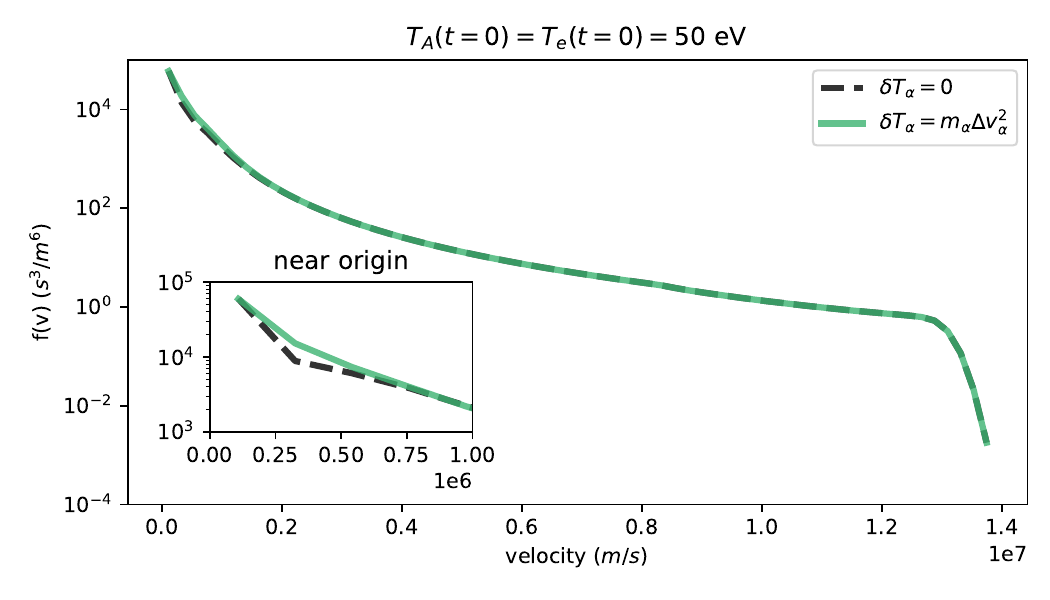}
    \caption{Comparison of steady-state \(\alpha\) distribution from an iFP simulation with (green solid line) and without (black dashed line) \(\alpha\) temperature flooring. The inset shows the comparison near the velocity space origin, demonstrating the smoothing effect of the $T$-floor.}
    \label{fig:alpha_T_floor_comp}
\end{figure}

Figure~\ref{fig:alpha_T_floor_comp} illustrates that introducing \( \delta T_\alpha \) as defined in Eq.~\eqref{eq:delta_T_floor_def} does not lead to any significant unphysical artifacts. Only minor deviations appear at low \(\alpha\) energies, where the numerical singularity at \( v \sim 0 \) is further smoothed.

\begin{figure}
    \centering
    \includegraphics[width=0.325\linewidth]{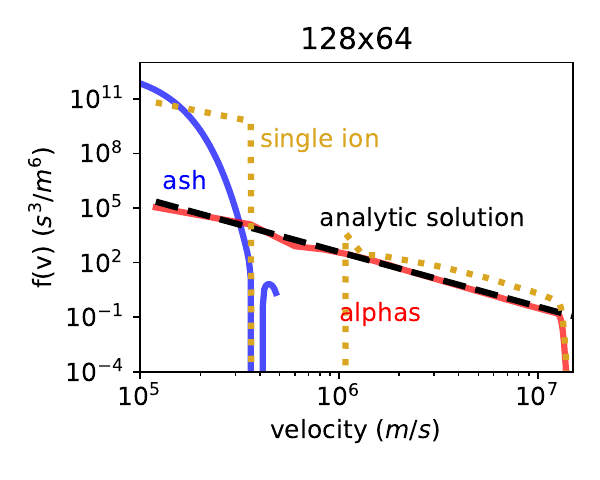}
    \includegraphics[width=0.325\linewidth]{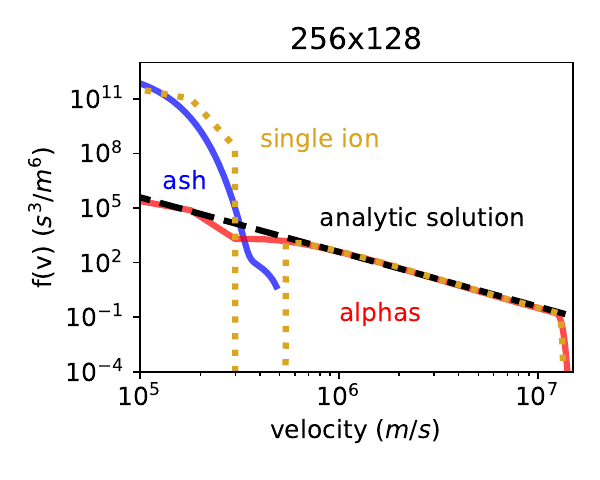} 
    \includegraphics[width=0.325\linewidth]{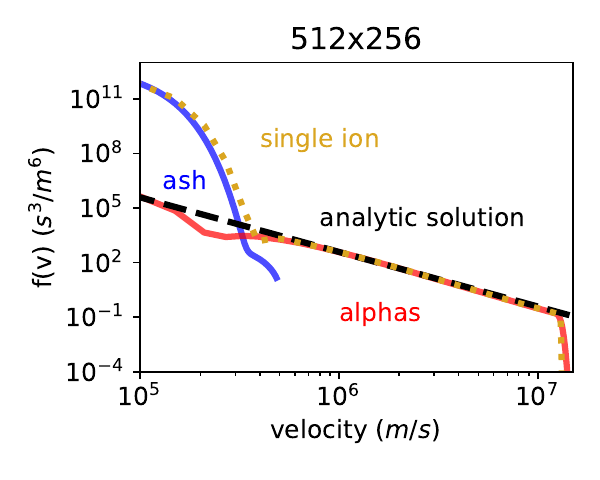} 
    \caption{Comparison of distribution function solution between a simulation with a single alpha/ash species (gold dotted line) with the two-grid approach for ash and alpha species (blue and red lines, respectively) as velocity space grid is refined from $N_\parallel \times N_\perp = 128 \times 64$ (left) to $256 \times 128$ (middle) to $512 \times 256$ (right).}
    \label{fig:one_ion_comp}
\end{figure}

The impact of the two-grid approach can be seen by observing that less velocity space resolution is necessary to recover the same ash and alpha distribution functions while avoiding significant unphysical features. Figure~\ref{fig:one_ion_comp} demonstrates a comparison between the analytical solution, the two-grid simulation, and a single-grid simulation as the velocity space grid resolution is increased. The initialization is the same as for Figure~\ref{fig:0d2v_slowing_down} but with $T_{A,0} = \qty{100}{eV}$. It is seen that the single-grid simulation requires much larger velocity resolutions to recover the correct distribution function and at lower resolutions is subject to numerical instabilities that lead to large negative distribution function values.
%
%
%
%
%
\subsection{0D2V $\alpha$-Heating of Electrons and Ash}
\label{subsec:0d2v_alpha_heating}
We compare the dynamic heating rate of fluid electrons due to a prescribed fixed source of $\alpha$-particles obtained from the iFP calculation with that from a semi-analytical model. Unlike the previous section, where a quasi-steady-state solution exists, this problem is inherently dynamic. This case serves as a quantitative test to demonstrate the robustness of the algorithm across a wide range of temperatures, beyond the limiting case of asymptotic scale separation, $v^*_{\alpha} \gg v_{th,A}$. The ODE model for the internal energy of fluid electrons is given by:
\begin{flalign}
    \label{eq:effective_temp_e}
    &
    \frac{3}{2} \frac{\partial n_e T_e}{\partial t} = 
    3 n_{He} \nu_{He,e} \left( T_{He} - T_{e} \right),
    &
\end{flalign}
where $T_{He}$ is the average temperature for the alpha and ash species combined (i.e., all $^4$He produced by DT fusion), $n_{He}$ is the combined alpha and ash density, and $\nu_{He,e} \equiv \nu_{\alpha e} = \nu_{Ae}$ is the $^4$He-electron collision frequency given by Eq.~\eqref{eqn:nu_ie} (the $\alpha$ and ash species share the same mass and charge), evaluated with the instantaneous $n_e$ and $T_e$. Because the electron energy exchange is weighted by the total $^4$He density, $n_{He}\nu_{He,e} = n_{\alpha}\nu_{\alpha e} + n_{A}\nu_{Ae}$, this is consistent with the sum of the $\alpha$ and ash contributions in Eq.~\eqref{eqn:fluid_e}. For the combined $^4$He species, the internal energy equation is:
\begin{flalign}
    \label{eq:alpha_temp}
    &
    \frac{3}{2} \frac{\partial n_{He} T_{He}}{\partial t} =
    \frac{m_{\alpha} v^2_{\alpha}}{2} S_{0} -
    3 n_{He} \nu_{He,e} \left( T_{He} - T_{e} \right),
    &
\end{flalign}
and the corresponding number density continuity equation is given by:
\begin{flalign}
    \label{eq:alpha_continuity}
    &
    \frac{\partial n_{He}}{\partial t} = S_{0}.
    &
\end{flalign}

For the initial conditions, we prescribe $n_{e,0} = \qty{2.52e24}{cm^{-3}}$, $n_{\alpha,0}=\qty{5.04e14}{cm^{-3}}$, $n_{A,0}=\qty{1.26e24}{cm^{-3}}$, $T_{e,0} = \qty{1.0}{keV}$, $T_{\alpha,0}=\qty{81.6}{keV}$, $T_{A,0} = \qty{1.0}{keV}$, $u_{\|,e} = 0$, $u_{\|,\alpha} = 0$, $u_{\|,A} = 0$, and a source strength of $S_{0} = \qty{2.16e23}{cm^{-3}/ns}$. The simulation is run over $t \in \left[ 0, \qty{0.17}{ns} \right]$.
\begin{figure}[h!]
    \centering
    \includegraphics[width=0.98\linewidth]{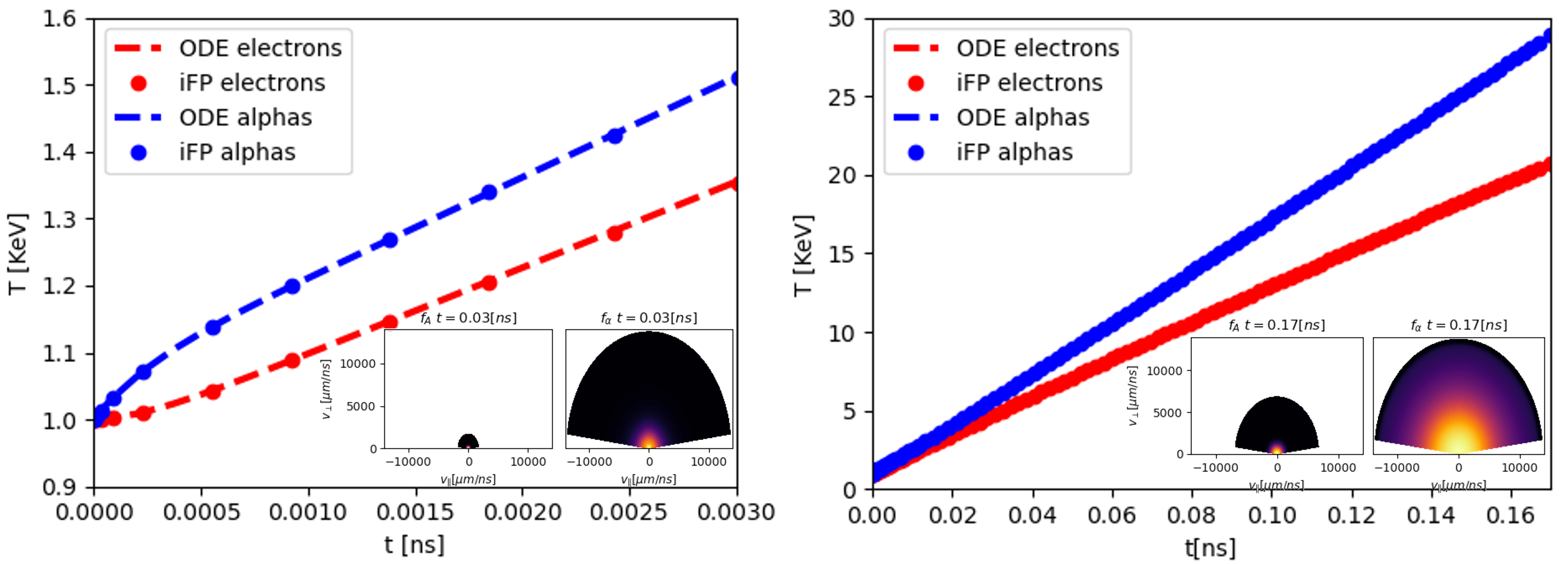}
    \caption{0D2V heating test: The blue circles represent the average temperature of the ash and $\alpha$-species combined, while the red circles correspond to the electron temperature from iFP simulations. The blue and red lines indicate the same quantities computed from the system of ODEs defined by Eqs.~\eqref{eq:effective_temp_e}, \eqref{eq:alpha_temp}, and \eqref{eq:alpha_continuity}. The left plot highlights the early-time behavior of the solution, while the right plot depicts the entire solution. Also shown are the ash and $\alpha$ distribution functions at early and late times.}
    \label{fig:0d2v_heating}
\end{figure}
The resulting electron and $\alpha$-particle temperatures, obtained from both the kinetic simulation and the ODE system defined by Eqs.~\eqref{eq:effective_temp_e}, \eqref{eq:alpha_temp}, and \eqref{eq:alpha_continuity}, are shown in Figure~\ref{fig:0d2v_heating}. To match with the ODE, the number-weighted average of the ash and $\alpha$ temperatures is used as output from the iFP simulation, while the ash and alpha number densities are summed:
\begin{flalign*}
    &
    T_{He} = \frac{n_A T_A + n_{\alpha} T_{\alpha}}{n_A + n_{\alpha}},
    &
\end{flalign*}

\begin{flalign*}
    &
    n_{He} = n_A + n_{\alpha}.
    &
\end{flalign*}
The close agreement between iFP results and the ODE system throughout the entire calculation---despite $\Omega_A$ becoming comparable to $\Omega_{\alpha}$---underscores the importance of a self-consistent formulation that satisfies detailed-balance relations.

%
%
\subsection{1D2V $\alpha$ Particle Slowing Down}
\label{subsec:1d2v_slowing_down}
To verify our implementation in a spatially inhomogeneous scenario, we consider the slowing down of \(\alpha\) particles within a spherical domain of radius \( R \). The extension of the algorithm to a spatially inhomogeneous setting is straightforward due to the 0D2V nature of the proposed method. Therefore, we refer interested readers to Ref.~\onlinecite{taitano_cpc_2021_ifp_code} for details on the 1D2V treatment. 

The 1D2V algorithm is compared to a semi-analytical solution for an isotropic alpha source slowing within a spherical domain of homogeneous background plasma. The details of this solution are given in Appendix~\ref{app:1d2v_solution}.

For boundary conditions in real space, we impose a homogeneous Neumann boundary condition on the electron temperature at \( r = 0 \) and a Dirichlet condition at \( r = R \) with \( T_e(r=R) = T_{e,0} \). For the ion distribution functions, we apply a specular reflective boundary at \( r = 0 \) and an inflow/outflow boundary at \( r = R \), where the inflow condition is prescribed by a Maxwellian distribution parameterized by \( n_{A,0} \), \( u_{\|,0} \), and \( T_{A,0} \). 

We consider a spherical domain \( r \in \left[ 0, R \right] \) with \( R = \qty{1122}{\um} \). The transformed velocity space is defined as \( \left( \widehat{w}_{\|} , \widehat{w}_{\perp} \right) \in \left[ -7.0, 7.0 \right] \times \left[ 0, 7 \right] \). The initial plasma is assumed to be static, with \( u_{\|,0} = 0 \) at hydrodynamic equilibrium, characterized by \( n_{A,0} = \qty{1e22}{cm^{-3}} \), \( n_{e,0} = \qty{2e22}{cm^{-3}} \), and \( T_{A,0} = T_{e,0} = \qty{1}{\keV} \). A fixed and uniform source of \(\alpha\)-particles is considered, given by \( S_0 = \qty{1.52e18}{cm^{-3}/ns} \), with an initial floor population of \( n_{\alpha,0} = \qty{1e16}{cm^{-3}} \) at \( T_{\alpha,0} = \qty{100}{\keV} \). The initial normalization speed is set to \( \left(v^*_{A,0},v^*_{\alpha,0}\right) = \left( \qty{219}{\um / \ns}, \qty{2190}{\um / \ns} \right) \), with a shift velocity of \( \left(u^*_{\|,A}, u^*_{\|,\alpha} \right) = \left( 0.0, 0.0 \right) \). A grid of \( N_r \times N_{\|} \times N_{\perp} = 192 \times 128 \times 64 \) is employed, and the simulation runs until \( t_{\max} \approx \qty{0.41}{\ns} \), when the solution has reached an approximate steady state. 

\begin{figure}[h!]
    \centering
    \includegraphics[scale=0.6]{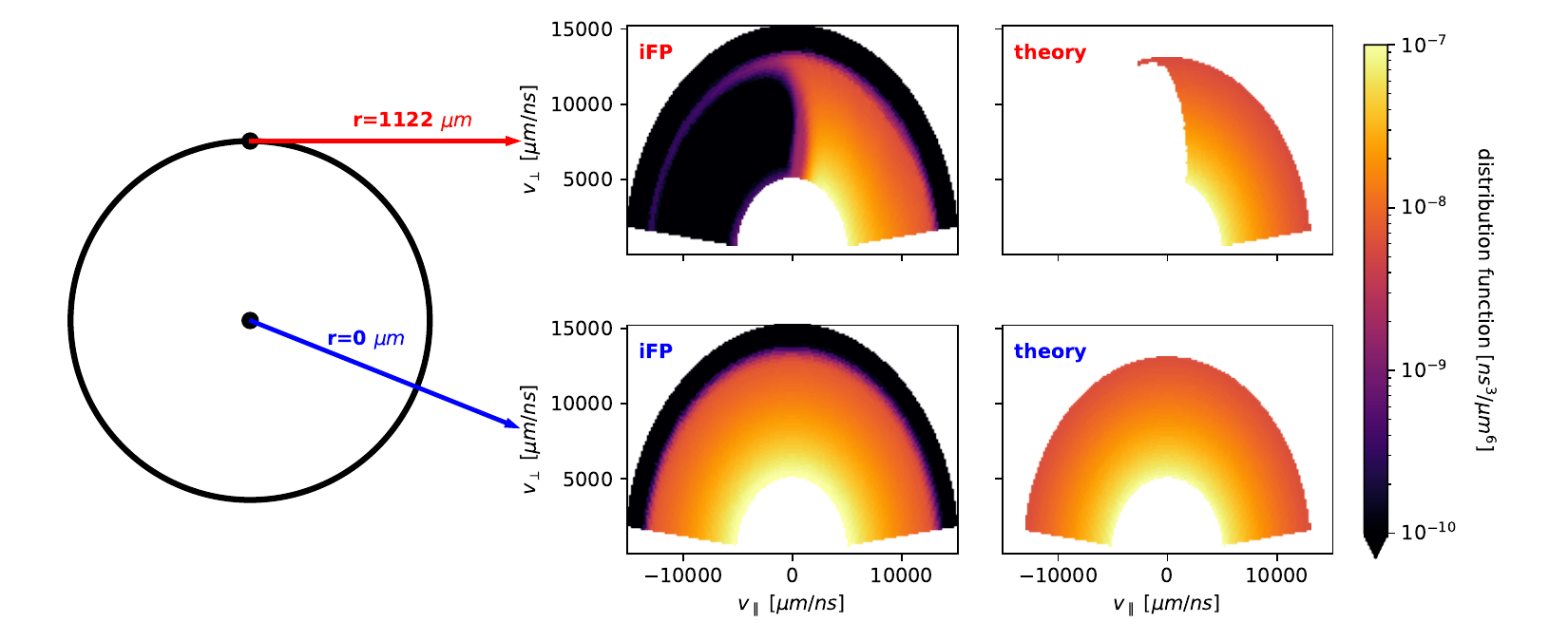}
    \caption{1D2V slowing down: The \(\alpha\)-particle distribution function at \( r \approx R \) (top) and \( r \approx 0 \) (bottom) from the iFP simulation (left) and analytical solution (right). The cylindrical coordinate system representation of the iFP solution has been mapped onto a polar coordinate system for comparison.}
    \label{fig:1d2v_slowing_down}
\end{figure}

Figure~\ref{fig:1d2v_slowing_down} presents the \(\alpha\)-particle distribution function at \( r=0 \) and \( r=R \).
As expected, the iFP distribution function accurately captures the phase-space profile of \(\alpha\)-particles slowing down against the fluid electrons. In addition, the effects of velocity-space diffusion absent in the semi-analytical solution cause visible smearing of the distribution function calculated by iFP.
%
%
%
%
%
\subsection{Convergence Study}
\label{subsec:convergence_study}
We demonstrate the formal order of convergence of the proposed method in a 1D2V spherical coordinate system with general velocity-space adaptivity for the ash population. Specifically, we consider a 1D2V spherical geometry, $r \in [0, R]$, where $R=\qty{2.67}{\mm}$ and with a transformed ion velocity-space domain of $\widehat{w}_{\|} \in [-7, 7]$ and $\widehat{w}_{\perp} \in [0, 7]$. The plasma is initially assumed to be static, $u_{\|,0} = 0$, and composed of $^4$He ions in thermal equilibrium at $T_0 = \qty{10}{\keV}$, with a Gaussian spatial profile given by
\begin{flalign}
    \label{eq:nhe4_profile}
    &
    n_{He4,0} = n_{floor,0} + n_{He4,0} e^{-\frac{(r - r_0)^2}{2\sigma^2_{He4}}}.
    &
\end{flalign}
Here, \(n_{floor,0} = \qty{4.2e19}{cm^{-3}}\) is a small floor density used to ensure numerical robustness, and 
\[\left(n_{He4,0}, r_0, \sigma_{He4}\right) = \left(\qty{4.2e21}{cm^{-3}}, \qty{1.34}{\mm}, \qty{0.291}{\mm}\right)\] 
denote the amplitude, mean, and standard deviation of the Gaussian profile, respectively. The energetic $\alpha$ particles are also initialized as static, $u_{\|,\alpha} = 0$, with $n_{\alpha,0} = 10^{-4} \times n_{He4,0}$ and $T_{\alpha} = \qty{120}{\keV}$. 
The source term has a Gaussian profile matching that of Eq.~\eqref{eq:nhe4_profile}, but without the floor density, and a peak source amplitude of $S_{0} = \qty{2.7e21}{\cm^{-3}/\ns}$. Electrons are treated self-consistently via quasi-neutrality, $n_e = -q^{-1}_e \sum_{\gamma \in {\cal G}} q_{\gamma} n_{\gamma}$, and are initialized to be in thermal equilibrium with the $^4$He ions, $T_{e,0} = T_{He4,0} = \qty{10}{\keV}$.  

\begin{figure}[t!]
    \centering
    \includegraphics[width=0.6\linewidth]{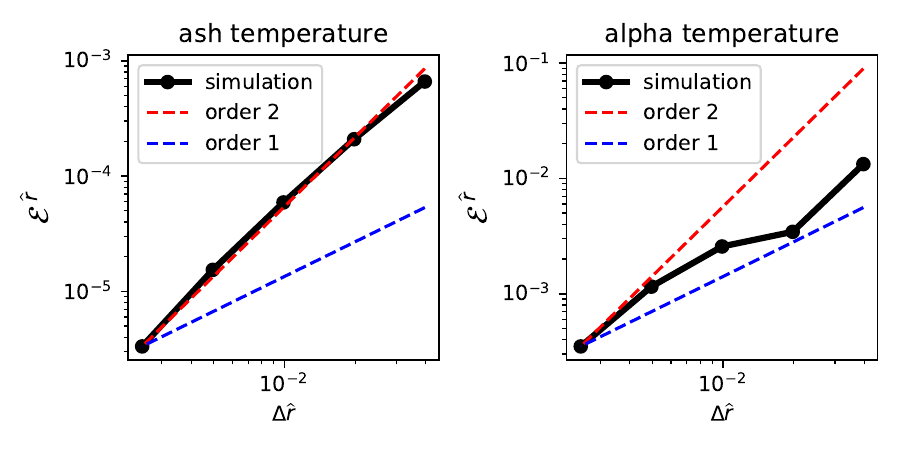}\\
    \includegraphics[width=0.6\linewidth]{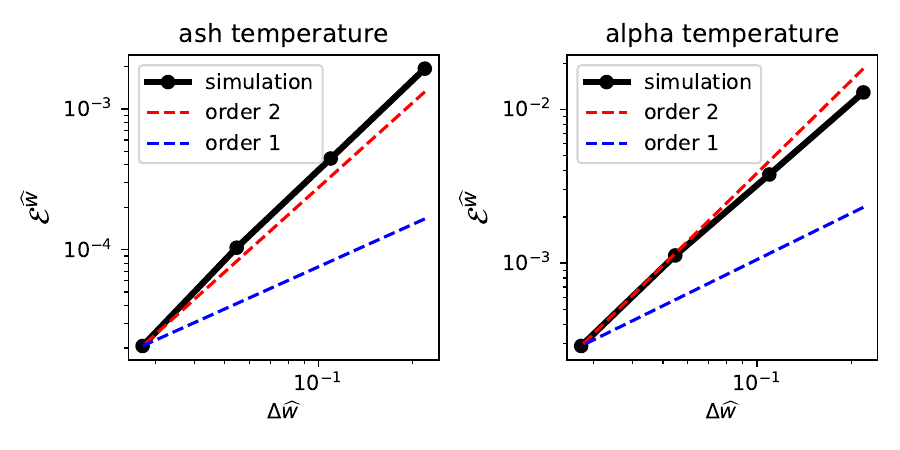}    
    \caption{Convergence study: Spatial grid refinement (top) and velocity-space grid refinement (bottom) for the temperature of ash (left) and alpha species (right). Here, $\Delta \widehat{r} = \Delta r / R$ is the domain normalized radial spatial grid resolution.}
    \label{fig:convergence_plot}
\end{figure}

For the convergence study, we compare temperature moments obtained using a coarse grid to those from a numerical reference solution computed with a fine grid.  In the velocity-space convergence study, the reference case uses $N_r = 24$, $N_\parallel \times N_\perp = 512 \times 256$, $t_{\max} = \qty{16}{\ps}$, and $\Delta t = \qty{0.13}{\ps}$. Convergence is evaluated on grids of $N_\parallel \times N_\perp = \{32\times16, 64\times32, 128\times64, 256\times128\}$.  In the position-space convergence study, the reference case uses $N_r = 768$, $N_\parallel \times N_\perp = 64 \times 32$, $t_{\max} = \qty{16}{\ps}$, and $\Delta t = \qty{0.13}{\ps}$, with convergence evaluated on grids of $N_r = \{24, 48, 96, 192, 384\}$. Figure~\ref{fig:convergence_plot} shows the $L_2$ norm of the temperature error for species $\gamma$, defined as:
\begin{flalign}
    \label{eqn:l2_norm_temperature}
    &
    \mathcal{E}_{\gamma} 
    = 
    \sqrt{
        \frac{
            \sum_{i=1}^{N_r} \Delta r\, r^2_i \left( T_{\gamma,i} - T^{\text{ref}}_{\gamma,i} \right)^2
        }
        {
            \sum_{i=1}^{N_{r,\text{ref}}} \Delta r_{\text{ref}}\, r^2_{\text{ref},i} \left(T^{\text{ref}}_{\gamma,i} \right)^2
        }
    }
    ,
    &
\end{flalign}
where the temperature moment for species $\gamma$ at position grid point $i$ is given by
\begin{flalign}
    \label{eq:temperature_moment}
    &
    T_{\gamma,i} = \frac{m_{\gamma}}{3} \frac{\left\langle \left| \mathbf{v} - \mathbf{u}_{\gamma,i} \right|^2, f_{\gamma,i} \right\rangle_{\delta v}}{n_{\gamma,i}}.
    &
\end{flalign}
For the position-space convergence study only, the reference solution is interpolated onto the coarse grid prior to computing the error.

As shown in Figure~\ref{fig:convergence_plot}, we observe asymptotic second-order convergence in all discrete norms. We note that, solely for the \textit{velocity-space grid convergence study}, a smoothing floor is added to the variance of fusion source definition (see Eq.~\eqref{eq:discrete_source_definition}), and defined as $\Delta v_{\alpha} := \sqrt{\Delta v_{\alpha}^2 + \Delta v_{\alpha, 256 \times 128}^2}$, where $\Delta v_{\alpha, 256 \times 128}$ is the grid spacing obtained from Eq.~\eqref{eq:grid_magnitude} using the $256\times128$ grid. This ensures sufficient smoothness of the reference solution to recover the formal second-order accuracy of the velocity-space discretization.
%
%
%
%
%
\subsection{1D2V Layered Spherical Implosion with DT-$\alpha$ Fusion Reaction}
\label{subsec:1d2v_layered_spherical_implosion}
We test the proposed algorithm on a system resembling a spherically imploding cryogenic layered ICF capsule, with an outer elastic moving wall acting as a piston. This test serves two purposes: (1) to evaluate the algorithm's performance in a realistic scenario where the $\alpha$-source depends self-consistently on the DT fuel density and temperature, and (2) to demonstrate the potential of the iFP implementation of the two-grid method as a surrogate for studying burn physics alongside fuel ion kinetic effects.

\begin{figure}[h]
    \centering
    \includegraphics[width=0.3\linewidth]{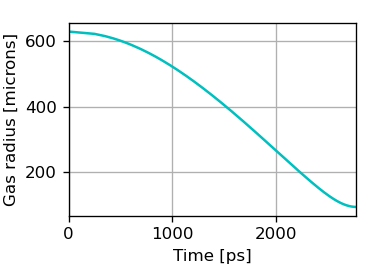}    
    \includegraphics[width=0.3\linewidth]{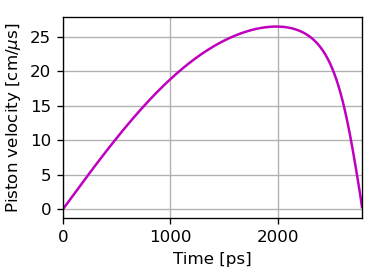}\\    
    \includegraphics[width=0.3\linewidth]{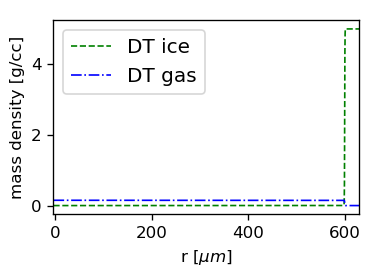}
    \includegraphics[width=0.3\linewidth]{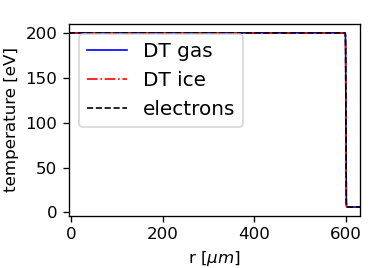}
    \caption{Layered spherical implosion: The outer DT vapor radius as a function of time (top left), the outer ice boundary velocity (top right), the initial mass densities (bottom left), and the initial temperatures (bottom right).}
    \label{fig:layered_implosion_ic_bc}
\end{figure}

To provide boundary conditions with the same characteristics of an ICF capsule, we adopt the implosion theory from Ref.~\onlinecite{montgomery_pop_2018_double_shell}. Assuming the outer radius of the cryogenic (ice) DT layer is driven inward by a constant pressure, the thin-shell approximation for the ice layer (or pusher) gives the total drive kinetic energy as
\begin{flalign}     
    \label{shell_KE}
    &
    T_\text{drive} \approx 2 \pi P_D R_0^3\left[\frac{2}{3}\left(\frac{C^3 - 1}{C^3}\right) - \left(\frac{\rho_0^\text{vapor}}{2\rho_0^\text{ice}}\right)(C^2 - 1)\right],
    &
\end{flalign}
where $P_D$ is the external (constant) drive pressure, $\rho_0^\text{ice}$ is the initial ice density, $\rho_0^\text{vapor}$ is the initial vapor density, and $C \equiv R_0/R(t)$ is the vapor convergence ratio, with $R_0$ and $R(t)$ representing the initial and time-dependent outer radii of the ice layer, respectively. Note that $R(t)$ formally denotes the radius at the pressure source (i.e., the outer radius of the ice layer). However, under the assumption that $d_p \ll R_0$ (where $d_p$ is the ice layer thickness), $R(t)$ approximately equals the inner radius of the vapor. The drive pressure is defined as
\begin{flalign}     
    \label{drive_pressure}
    &
    P_D = \frac{3}{2}\left(\frac{d_p}{R_0}\right)\rho_0^\text{ice}U_\text{max}^2,
    &    
\end{flalign}
where $U_\text{max}$ is the prescribed maximum pusher velocity. By integrating Eq.~\eqref{shell_KE}, we obtain $R(t)$ and $U(t) = \dot{R}$. The motion of iFP's piston boundary condition---defined at the outer radius of the ice layer---is then governed by $R(t)$ and $U(t)$. For our problem, we choose $C = 6.73$, $R_0=600\,\mu$m, $P_D = 0.3375$~Gbar, $\rho_0^\text{vapor} = 0.15$~g/cm$^3$, $\rho_0^\text{ice} = 5$~g/cm$^3$, $U_\text{max} = 30$~cm/$\mu$s, and $d_p = 30\,\mu$m. Based on total mass conservation, this setup should yield an areal density greater than $\rho R \approx 0.4$~g/cm$^2$, exceeding the ignition threshold based on the Atzeni-Meyer-ter-Vehn criterion\cite{atzeni_2004_book_pif}. We note, however, that since we neglect radiation physics and do not allow conduction loss across the piston wall, the ignition condition is relaxed significantly in this study. The initial condition and outer radius evolution over time are shown in Fig.~\ref{fig:layered_implosion_ic_bc}.

We consider a computational domain of $\left(\xi , \widehat{w}_{\|}, \widehat{w}_{\perp}\right) \in \left[ 0, 1 \right]\times \left[ -7,7\right] \times \left[0, 7 \right]$ with a grid resolution of $N_{\xi} \times N_{\|} \times N_{\perp} = 384 \times 64 \times 32$. Here, $\xi$ is the logical radial coordinate (to be discussed shortly), and $N_{\xi}$ represents the number of grid points in that dimension. The ion species considered are deuterium, tritium, ash, and $\alpha$-particles. A Neumann boundary condition is applied to the electron temperature at $\xi = 0$ and $\xi = 1$, while for the ion distribution function, a specular reflective condition is imposed at $\xi = 0$, and an elastic reflective moving wall boundary condition at $\xi = 1$; see Ref.~\onlinecite{taitano_cpc_2021_ifp_code} for further details.

To facilitate practical calculations, we employ several additional numerical techniques specific to this problem. This paragraph is provided for reproducibility, and readers may skip it if desired. Initially, $\alpha$-particles are not dynamically evolved until an empirically chosen time of $t = 2.56$~ns, when thermonuclear fusion reactions become relevant. This avoids the need to resolve fast $\alpha$-particle advection timescales early on in the calculation. To capture the evolving spatial structures efficiently, we employ a nonlinearly stabilized moving-grid strategy based on error equidistribution and a monitor function that adapts the grid near sharp gradients at each timestep, as detailed in Refs.~\onlinecite{taitano_cpc_2021_ifp_code, taitano_cpc_2021_vth_u_shift}. The solution is evolved on a uniform logical computational domain, $\xi$, and mapped to physical space through the grid Jacobian. Contributions from $\alpha$- and ash-species are deliberately excluded from the monitor function to prevent brittle behavior in the adaptive algorithm due to source-sink effects. To address the wide range of timescales in the implosion process, we heuristically adapt the timestep following the approach in Ref.~\onlinecite{taitano_cpc_2021_ifp_code}.

\begin{figure}[h!]
    \centering
    \includegraphics[width=0.96\linewidth]{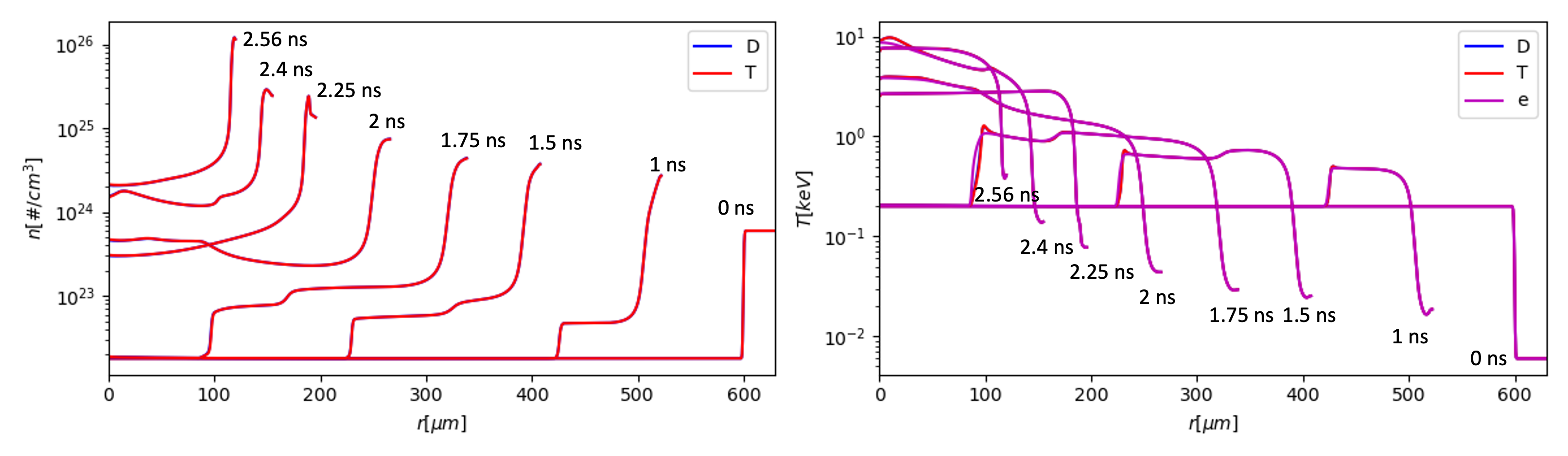}
    \includegraphics[width=0.96\linewidth]{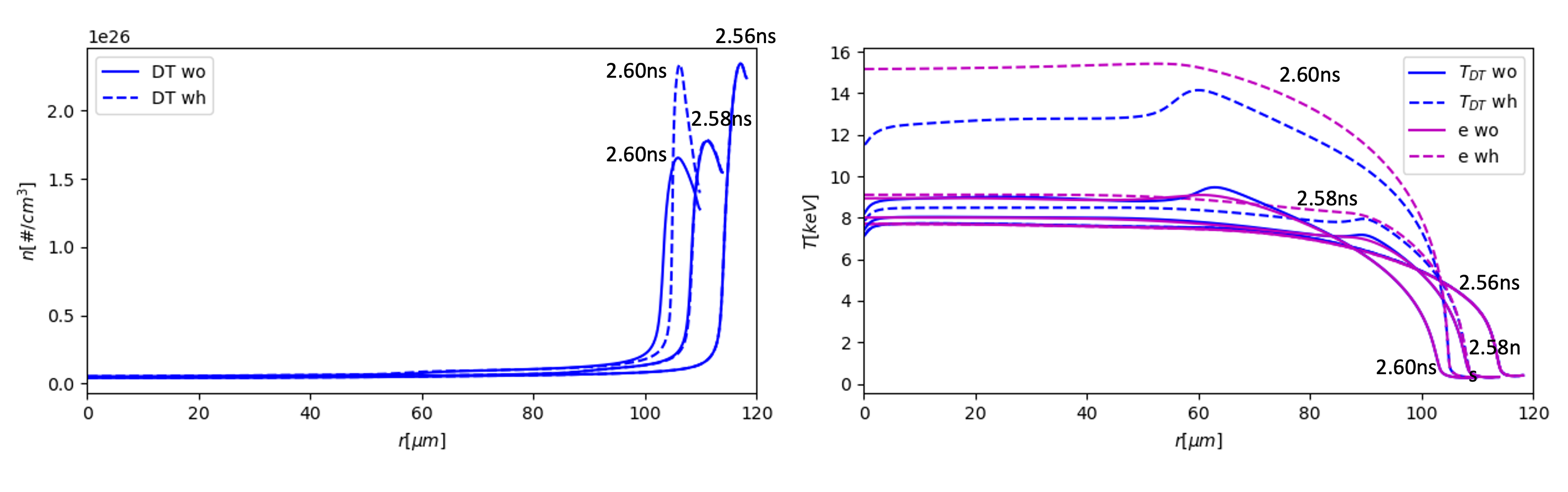}
    \caption{1D2V layered spherical implosion: The number density of DT ions (top left) and temperatures with electrons (top right) at early times. The second row compares the number density (left) and temperature (right) with (dashed line) and without (solid line) fusion reactions at later times.}
    \label{fig:1d2v_implosion_simulation_moments}
\end{figure}

\begin{figure}[h!]
    \centering
    \includegraphics[width=0.96\linewidth]{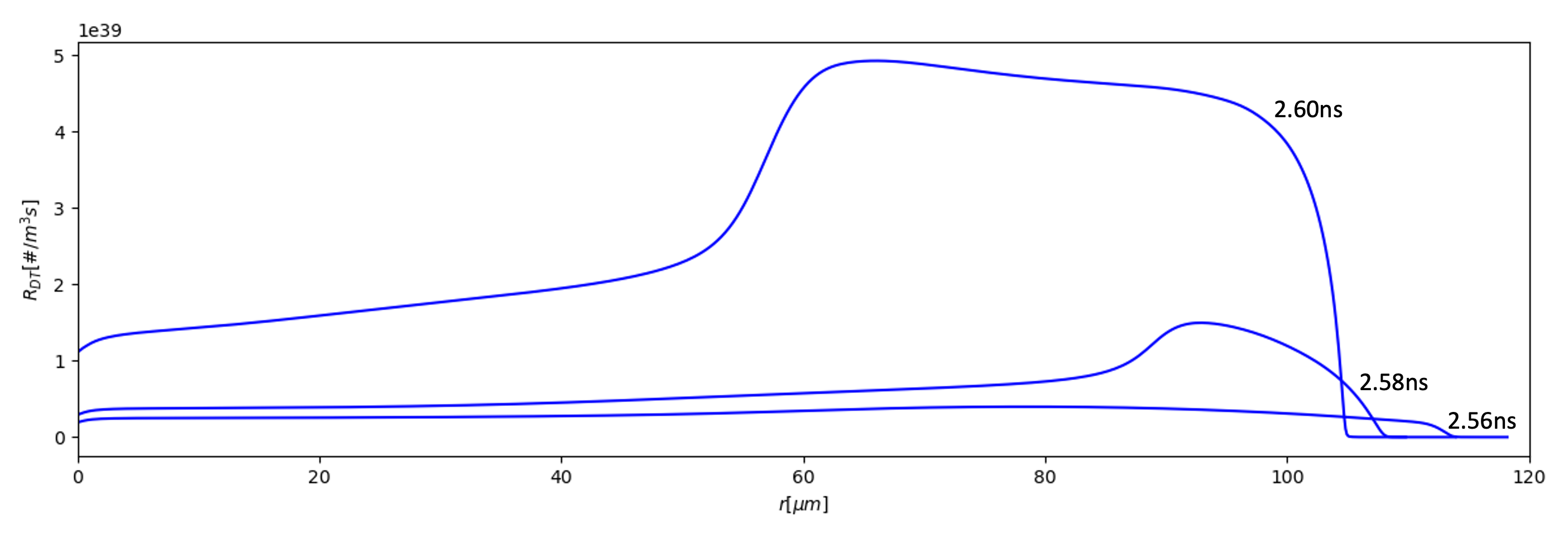}\\
    \includegraphics[width=0.7\linewidth]{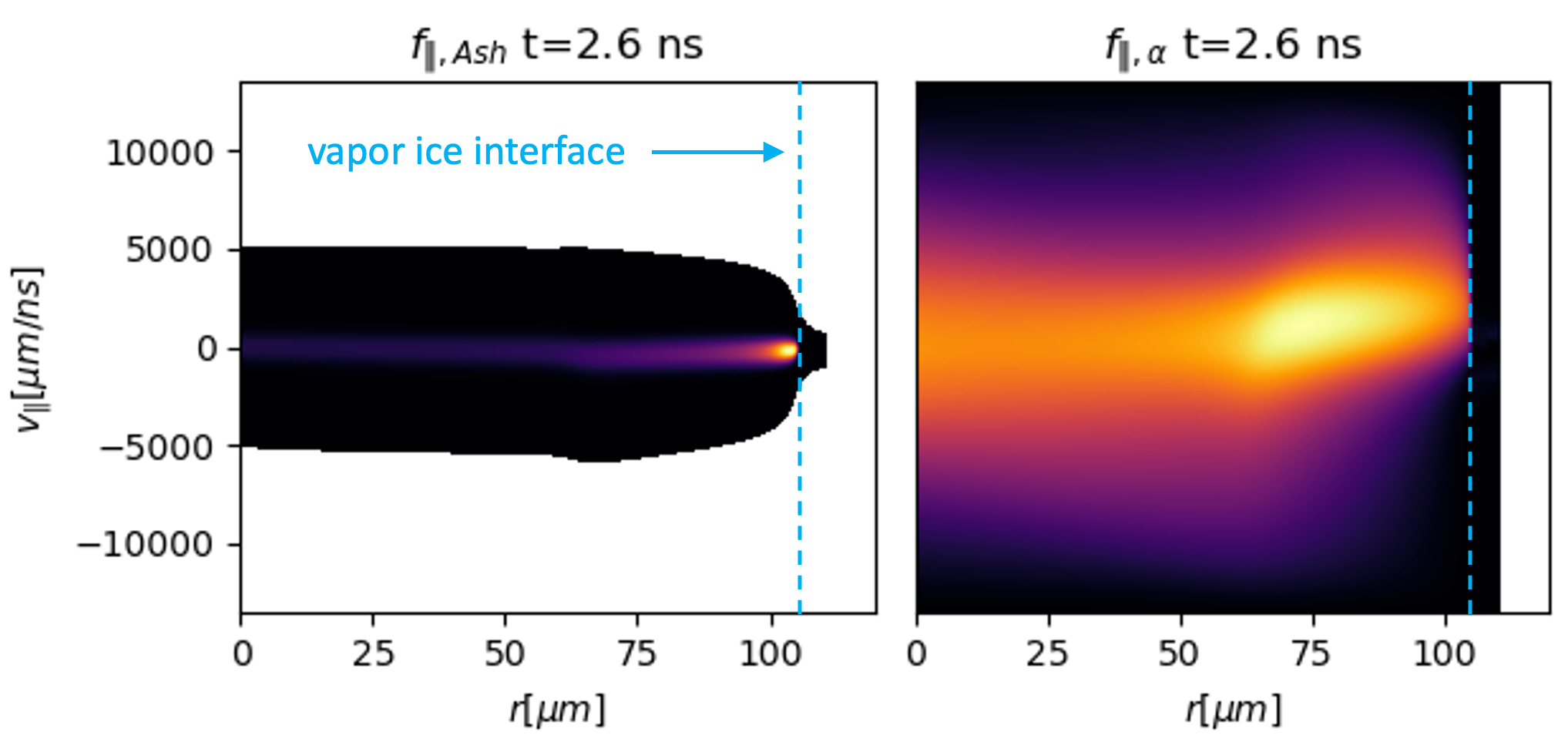}\\
    \caption{1D2V layered spherical implosion: The DT-$\alpha$ reactivity at late times (top), along with the parallel marginal distribution functions for ash (left) and $\alpha$ particles (right) at $t = 2.60$~ns (bottom).}
    \label{fig:1d2v_implosion_simulation_reactivity_pdf}
\end{figure}

We now discuss the numerical results. Figure~\ref{fig:1d2v_implosion_simulation_moments} presents the early and late-time density and temperature profiles of DT fuel ions and electrons.
Up to $t \sim 2.56$~ns, the implosion compresses and heats the DT plasma to thermonuclear conditions. When fusion reactions are activated, $\alpha$-particles rapidly heat the vapor region, triggering a self-sustained burn. The subsequent pressure increase leads to outward compression of the ice region around $t \sim 2.60$~ns. Figure~\ref{fig:1d2v_implosion_simulation_reactivity_pdf} presents the late-time DT-$\alpha$ reactivity and the corresponding parallel marginal distribution functions, $f_{\|} = 2\pi \int dv_{\perp} v_{\perp} f$, highlighting $\alpha$'s collisionally slowing down near the ice-vapor interface. 
As can be seen, most of the ash particles generated via $\alpha$-particle slowing-down accumulate near the ice-vapor interface, while they are nearly absent in the ice itself. This behavior arises due to the intense reactivity within the ablated ice region near the interface, where large amounts of $\alpha$-particles are produced but rapidly thermalize due to extreme collisionality. Consequently, the energy deposition is highly localized near the interface, leading to significant heating of the vapor region while leaving the ice layer largely unaffected.
%
%
%
%
%
\section{Conclusions and Future Directions}
\label{sec:conclusions}
We have proposed an Eulerian hybrid fluid electrons and ion Vlasov-Fokker-Planck algorithm (for all ion species including fuel and \(\alpha\) particles) that can robustly model multiscale burning plasma conditions relevant to igniting ICF capsules. We have formulated a multiscale, consistent and conservative, two-grid algorithm to model the dynamics of fusion-born energetic and thermal ion populations. In our approach, velocity coordinates are scaled for each population according to their characteristic speeds (energetic or thermal), enabling the accurate resolution of both energetic and thermal scales. The interaction between these two populations is mediated by collisions and numerical sink/source terms that preserve detailed balance as well as total mass, momentum, and energy of the two populations. The velocity-space-dependent sink/source rates are derived from a physics-based reduced-order model.

We have demonstrated the advantages and robustness of the new algorithm on problems of increasing complexity, ranging from 0D2V to a 1D2V spherical implosion problem, which emulates layered cryogenic capsules in ICF experiments (e.g., those at the Omega and NIF facilities\cite{abu_shawareb_2022_prl_ignition,haines_2024_pop_ignition_simulation}). Looking ahead, two points merit attention: 1) credible burning-plasma calculations require coupling with radiation transport, which has been implemented into iFP and will be documented in a follow-up manuscript; and 2) although we focus here on the source/sink interactions in energetic and thermal species, an effective implicit time-integration strategy is critical to handle the extremely stiff collision-time scales in dense DT ice layers, which will be documented in detail in the future. Future work will also consider on-the-fly evaluation of the self-consistent five-dimensional fusion reactivity integral, possibly accelerated with modern low-rank tensor decomposition methods for efficiency\cite{oseledets_siam_jsc_2011_ttd}.

%
%
%
%
%
\section*{Acknowledgement}
\label{acknowledgement}

B.L.R. was supported by the DOE Stockpile Stewardship Graduate Fellowship program.
W.T.T. was supported in part by the LANL ASC Thermonuclear Burn Initiative (2017--2020), the LANL Office of Experimental Sciences (2023--2024), and the DOE Office of Applied Scientific Computing Research (ASCR) through the Mathematical Multifaceted Integrated Capability Centers program (2025). 
A.N.S., L.C., H.R.H. and S.E.A. were supported by the LANL Directed Research and Development program. 
This work was also supported by the Institutional Computing program at Los Alamos National Laboratory and was performed under the auspices of the National Nuclear Security Administration of the U.S. Department of Energy at Los Alamos National Laboratory, managed by Triad National Security, LLC under Contract No. 89233218CNA000001.

%
%
%
%
%
\appendix
\section{Conservation of moments with floor temperature}
\label{app:moment_cons_thybridfloor}

By taking the energy moment of Eq.~\eqref{eqn:ie_fp_modified}, we demonstrate that the energy contribution from the floor temperature term cancels out with the corresponding term on the right-hand side of Eq.~\eqref{eqn:fluid_e_modified}. Integrating by parts once, we obtain:
\begin{flalign}
    \label{eqn:energ_mom_fluid_e_modified}
    &
    m_{\alpha} \left< \frac{v^2}{2} , C^*_{\alpha e} \right>_v 
    =
    -\nu_{\alpha e} 
    \left[ 
        \left(T_e + \delta T_{\alpha}\right) \left< {\bf v} , \nabla_v f_{\alpha} \right>_v 
        +
        m_{\alpha}\left< {\bf v} , \left({\bf v} - {\bf u}_{\alpha} \right) f_{\alpha} \right>_v
    \right].
    &
\end{flalign}
Evaluating the remaining moments, $\left< {\bf v} , \nabla_v f_{\alpha} \right>_v = -3 n_{\alpha}$ (by a second integration by parts) and $m_{\alpha}\left< {\bf v} , \left({\bf v} - {\bf u}_{\alpha} \right) f_{\alpha} \right>_v = m_{\alpha}\left< \left|{\bf v} - {\bf u}_{\alpha}\right|^2 , f_{\alpha} \right>_v = 3 n_{\alpha} T_{\alpha}$, we obtain:
\begin{flalign}
    \label{eqn:energ_mom_fluid_e_modified_2}
    &
    m_{\alpha} \left< \frac{v^2}{2} , C^*_{\alpha e} \right>_v 
    =
    \nu_{\alpha e}
    \left[ 
        3 n_{\alpha} \left( T_e + \delta T_{\alpha} - T_{\alpha}  \right) 
    \right].
    &
\end{flalign}
Since \( n_\alpha \nu_{\alpha e} = n_e \nu_{e \alpha} \) by definition, energy balance between the energetic population and fluid electrons is preserved:
\begin{flalign}
    \label{eq:energy_sum_alpha_electron}
    &
    \frac{3}{2}\frac{\partial}{\partial t}\left( n_e T_e \right) + m_{\alpha} \left< \frac{v^2}{2} , C^*_{\alpha e} \right>_v 
    = 
    3\sum_{\gamma \in {\cal G}^{\backslash \alpha}} n_e \nu_{e\gamma} \left( T_{\gamma} - T_e \right).
    &
\end{flalign}
Note that momentum conservation is not affected by the new term, as can be trivially shown by taking the first velocity moment of the term, which vanishes.

\section{Details of moment-conserving source term}
\label{app:moment_cons_source}

The additive correction term is defined as:
\begin{flalign}
    \label{eq:additive_correction_def}
    &
    \delta S_{A\leftarrow\alpha,j,k} = S_{A\leftarrow \alpha,j,k} \sum^2_{l=0} C_l \psi_{l,j,k},
    &
\end{flalign}
where $\delta S_{A\leftarrow\alpha,j,k}$ ensures balance of total mass, momentum, and energy between $S_{\alpha\rightarrow A}$ and $S_{A\leftarrow \alpha}$. The coefficient $C_l$ is the $l^{th}$ projection coefficient, while $\psi_{l,j,k} = \psi_l\left(\widehat{\bf w}_{j,k} \right)$ represents the projection function, chosen as:
\begin{flalign}
    \label{eq:proj_func_def}
    &
    \boldsymbol{\psi} = \left\{ \psi_0, \psi_1, \psi_2 \right\} = 
    \left\{1, 
    w_{\|,A\leftarrow\alpha}, 
    w^2_{A\leftarrow\alpha} \right\}.
    &
\end{flalign}
The velocity components are defined as:
\begin{flalign}
    \label{eq:proj_func_1}
    &
    w_{\|,A\leftarrow\alpha} = \widehat{w}_{\|}v^*_{A} + u^*_{\|,A} - u_{\|,A\leftarrow\alpha},
    &
\end{flalign}
\begin{flalign}
    \label{eq:proj_func_2}
    &
    w^2_{A\leftarrow\alpha} = w^2_{\|,A\leftarrow\alpha} + v^2_{\perp,A},
    &
\end{flalign}
\begin{flalign}
    \label{eq:proj_func_3}
    &
    u_{\|,A\leftarrow\alpha} = \frac{\left<v_{\|,A},S_{A\leftarrow\alpha} \right>_{\delta v_{A}}}{\left<1 , S_{A\leftarrow\alpha} \right>_{\delta v_{A}}}.
    &
\end{flalign}

The projection coefficients, $\boldsymbol{C} = \left\{C_0, C_1, C_2 \right\}$, are determined by solving the following optimization problem:
\begin{flalign}
    \label{eq:minimization_problem_cons_proj}
    &
    \min_{{\bf C},\boldsymbol{\lambda}}{\bf C}^T \cdot {\cal M} \cdot {\bf C} 
    - 
    \boldsymbol{\lambda} \cdot \left[ \left<\boldsymbol{\psi} , S_{A\leftarrow\alpha} + \delta S_{A\leftarrow\alpha}\right>_{\delta v_A} - \left< \boldsymbol{\psi} , S_{\alpha\rightarrow A} \right>_{\delta v_{\alpha}} \right].
    &
\end{flalign}
Here, ${\cal M} \in \mathbb{R}^{3\times3}$ is a diagonal weight matrix with entries:
\begin{flalign*}
    &
    {\cal M}_{00} = 1,
    &
\end{flalign*}
\begin{flalign*}
    &
    {\cal M}_{11} = \left| W_{\|,A\leftarrow\alpha} \right|_{\infty},
    &
\end{flalign*}
\begin{flalign*}
    &
    {\cal M}_{22} = \left| W_{\|,A\leftarrow\alpha}\right|^2_{\infty} + \left|V_{\perp,A}\right|^2_{\infty},
    &
\end{flalign*}
which penalize the linear and quadratic contributions more heavily, particularly for larger velocities. Here, 
\begin{flalign}
    &
    W_{\|,A\leftarrow\alpha} = \widehat{W}_{\|}v^*_{A} + {\bf e}_{\|} \left( u^*_{\|,A} - u_{\|,A\leftarrow\alpha}\right),
    &
\end{flalign}
$\boldsymbol{\lambda} 
\in \mathbb{R}^3$ is the vector of Lagrange multipliers for the constraints. Finally, we note that the introduction of $u_{\|,A\leftarrow\alpha}$ tailors the projection basis to ensure that the modification of the source, $S_{A\leftarrow \alpha}$, is centered around the mean velocity defined by the pre-corrected source, ${\bf S}_{A\leftarrow\alpha}$. We note that, although the detailed balance of $S_{\alpha\rightarrow A}$ and $S_{A\leftarrow\alpha}$ leads to preservation of infinite number of moments (e.g., entropy), we restrict ourselves in this study to preserve only the first three. They are the relevant quantities for capturing the important macroscopic hydrodynamic behaviors such as shock speed, which are determined by mass, momentum, and energy balance relations (e.g., the Rankine-Hugoniot conditions). 
\section{Chang-Cooper discretization of collisional flux}
\label{app:chang_cooper}
Consider the evaluation of the ion-fluid-electron collision operator at cell $\left( j, k \right)$:

\begin{flalign}
    &
    C_{\gamma e,j,k} \approx 
    \nu_{\gamma e} 
    \left[ 
        \frac{\widehat{J}_{\|, j+\frac{1}{2},k} - \widehat{J}_{\|,j-\frac{1}{2},k}}{\Delta v_{\|}} 
        + 
        \frac{v_{\perp,k+\frac{1}{2}} \widehat{J}_{\perp,j,k+\frac{1}{2}} - v_{\perp,k - \frac{1}{2}} \widehat{J}_{\perp,j,k -\frac{1}{2}}}{v_{\perp,k}\Delta v_{\perp}}
    \right].
    &
\end{flalign}
Here, $\widehat{J}$ denotes the numerical reconstruction of the collisional fluxes at the cell interfaces. The Chang-Cooper discretization imposes the following fluxes:
\begin{flalign}
    \label{eq:chang_cooper_flux}
    &
    \widehat{J}_{\|,j+\frac{1}{2},k} 
    = 
    \frac{T_e + \delta T_{\gamma}}{m_{\gamma}} \frac{f_{\gamma,j+1,k} - f_{\gamma,j,k}}{\Delta v_{\|}} 
    + 
    \left(v_{\|,j+\frac{1}{2}} - u_{\|,\gamma} \right) \left[ \theta_{\|,j+\frac{1}{2}}f_{\gamma,j,k} + \left( 1 - \theta_{\|,j+\frac{1}{2}} \right) f_{\gamma,j+1,k} \right].
    &
\end{flalign}
Here, $\delta T_{\gamma} := \delta T_{\alpha} \delta_{\gamma \alpha}$, where $\delta_{\gamma\alpha}$ is the Kronecker delta function:
\begin{flalign}
    &
    \delta_{\gamma\alpha} = 
    \begin{cases}
        1, & \text{if } \gamma = \alpha, \\
        0, & \text{otherwise}.
    \end{cases}
    &
\end{flalign}
The Chang-Cooper interpolation weight, $\theta_{\|,j + \frac{1}{2}}$, is defined as:
\begin{flalign}
    \label{eq:chang_cooper_weight}
    &
    \theta_{\|,j+\frac{1}{2}} 
    = 
    -\frac{\frac{\left( T_e + \delta T_{\gamma} \right) \left(f^M_{\gamma,j+1,k} - f^M_{\gamma,j,k} \right)}{m_{\gamma} w_{\|,j+\frac{1}{2}} \Delta v_{\|}} + f^M_{\gamma,j+1,k}  }
    {f^M_{\gamma,j,k} - f^{M}_{\gamma,j+1,k}}.
    &
\end{flalign}
A similar procedure is applied to the perpendicular components. The goal of the Chang-Cooper discretization is to drive the solution toward an equilibrium distribution, $f^M_{\gamma}$, given by:
\begin{flalign}
    \label{eq:numerical_equilibrium_chang_cooper}
    &
    f^M_{\gamma,j,k} \propto \exp \left[ - \frac{m_{\gamma} \left( w^2_{\|,\gamma,j} + v^2_{\perp,k} \right)}{2\left( T_e + \delta T_{\gamma} \right)} \right],
    &
\end{flalign}
and to maintain this equilibrium if initially present. Thus, when all species are in thermal and drift equilibrium, the distribution function relaxes to a Maxwellian defined in terms of the common temperature and drift velocity.

\section{Semi-analytical 1D2V stopping solution}
\label{app:1d2v_solution}

\begin{figure}[h]
    \centering
    \includegraphics[scale=0.6]{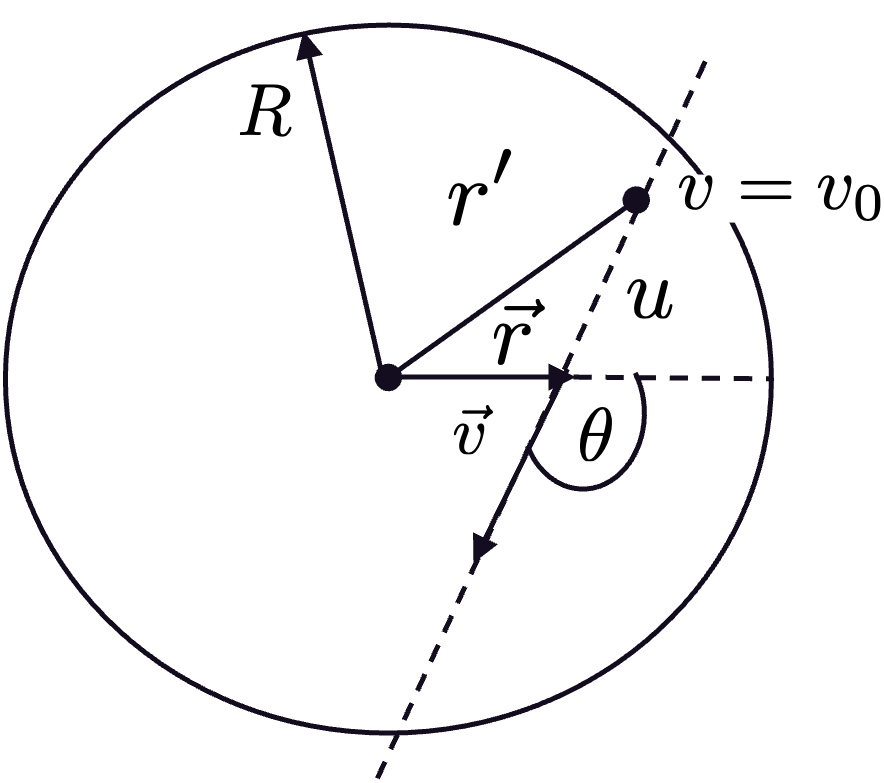}
    \caption{1D2V slowing down: Relationship between the various variables for \(\alpha\)-particle slowing down in spherical geometry.}
    \label{fig:spherical_diagram}
\end{figure}

The semi-analytical solution for the distribution of \(\alpha\)-particles slowing down against a fixed background of ash and electrons is derived by solving the particle characteristics for a given slowing down rate. From Ref.~\onlinecite{gaffey_energetic_ion_dist_1976}, if diffusion in velocity space is ignored and an isotropic alpha source at fixed velocity is given, the Vlasov-Fokker-Planck equation reduces to:

\begin{flalign}
    \label{eqn:vfp_multid}
    &
   \partial_t f_{\alpha} + \mathbf{v} \cdot \nabla_\mathbf{x} f_{\alpha} = \frac{1}{\tau_s v^2} \frac{\partial}{\partial v} \left[(v^3 + v_c^3) f_\alpha \right] + \frac{S_{\alpha}(\mathbf{x})\delta(v_\alpha-v)}{4\pi v_\alpha^2},
    &
\end{flalign}
where $v = \lVert \mathbf{v} \rVert$ and the $4\pi v_\alpha^2$ in the source term is from normalizing when integrating over velocity space. Expanding the gradient, we obtain the following form:
\begin{flalign}
    \label{eqn:vfp_multid_char}
    &
    \partial_t f_{\alpha} + \mathbf{v} \cdot \nabla_\mathbf{x} f_{\alpha} - \frac{v^3 + v_c^3}{v^2 \tau_s} \frac{\partial f_\alpha}{\partial v} = \frac{3}{\tau_s} f_\alpha + S_{\alpha}
    &
\end{flalign}
The left hand side becomes a total derivative along the characteristics parameterized by $t'=t$ defined by the equations:
\begin{flalign}
    \label{eqn:char_eqn_1}
    &
    \frac{dt}{dt'} = 1
    &
\end{flalign}
\begin{flalign}
    \label{eqn:char_eqn_2}
    &
    \frac{d\mathbf{x}}{dt'} = \mathbf{v}
    &
\end{flalign}
\begin{flalign}
    \label{eqn:char_eqn_3}
    &
    \frac{dv}{dt'} = - \frac{v^3 + v_c^3}{v^2 \tau_s}
    &
\end{flalign}
In other words, the characteristics are straight trajectories with monotonically decreasing velocity. Eq.~\eqref{eqn:char_eqn_3} can be used to find $t$ as a function of $v$ along each characteristic, with $t=0$ corresponding to $v=v_\alpha$. 
Recasting the left hand side as the total derivative along the characteristics $df_\alpha/dt$ allows Eq.~\eqref{eqn:vfp_multid_char} to admit an integrating factor of $H = c(v^3 + v_c^3)/(v_\alpha^3 + v_c^3)$, where $c$ is an arbitrary integration constant. Multiplying both sides by $H$ and moving the $f_\alpha$ term to the left hand side obtains the total derivative $(d/dt)(Hf_\alpha)$, which can be integrated. We enforce $f_\alpha(v>v_\alpha)=f_\alpha(t<0)=0$ since there is no source for any such particles. Leveraging behavior of the Dirac delta composed with a function, we find:
\begin{flalign}
    &
    \label{eqn:1d2v_fH_eqn}
    (H f_\alpha)|_t = \int_{0^-}^t H \frac{S_{\alpha}(\mathbf{x}(t'))\delta(v_\alpha - v(t'))}{4\pi v_\alpha^2} dt' = \int_{0^-}^t H \frac{S_{\alpha}(\mathbf{x}(t'))\delta(t')}{4\pi v_\alpha^2 \lvert \frac{dv}{dt'}(t=0)\rvert } dt'
    &
\end{flalign}
The sifting property of the Dirac distribution finally yields the semi-analytical form for the distribution function:
\begin{flalign}
    &
    \label{eqn:1d2v_semi_analytical_solution}
    f_\alpha(\mathbf{x}(t),\mathbf{v}) =  \frac{S_{\alpha}(\mathbf{x}(0))\tau_s}{4\pi (v^3 + v_c^3)}
    &
\end{flalign}
In the test case discussed in section~\ref{subsec:1d2v_slowing_down}, the source is homogeneous and constant in time over a sphere, and 0 elsewhere. Thus, $f_\alpha(\mathbf{x}(t),\mathbf{v})$ is a function only of $r= \lVert \mathbf{x}(t) \rVert$, $v$, and particle pitch angle \( \mu = \cos\theta \). This leads to the solution:
\begin{flalign}
    \label{eq:1d2v_semi_analytical_solution}
    &
    S_{\alpha}(\mathbf{x}(0)) = \left\{
        \begin{array}{ccc}
            S_0     &       \text{if}         &       r'<R \\
            0       &       \text{otherwise}
        \end{array}
    \right\}
\end{flalign}
Here, \( r' = \sqrt{r^2 + u^2 - 2 \mu r u} \), where \( \mu = \cos\theta \) is the particle pitch angle, and \( u \) represents the \(\alpha\)-particle slowing-down distance in velocity space (from the initial birth velocity \( v_{\alpha} \) to a target velocity \( v \)). This is found by multiplying Eq.~\eqref{eqn:char_eqn_3} by $v^{-1}$ to obtain the derivative of velocity with respect to path length and integrating to find:
\begin{flalign}
    \label{eq:slowing_down_distance}
    &
    u\left(v_{\alpha},v \right) = \frac{1}{\nu_s} \left\{ v_{\alpha} - v - \frac{v_c}{\sqrt{3}} 
        \left[
            \tan^{-1}\left( \frac{2v_{\alpha} - v_c}{\sqrt{3} v_c} \right) - 
            \tan^{-1}\left( \frac{2v - v_c}{\sqrt{3} v_c} \right)
        \right] 
        - \frac{v_c}{6} \log \left[ \left( \frac{v_{\alpha} + v_c}{v + v_c} \right)^2 \frac{v^2 - v v_c + v^2_c}{v^2_{\alpha} - v_c v_{\alpha} + v^2_c} \right]
    \right\}
    &
\end{flalign}
where \( \nu_s = \tau^{-1}_s \) denotes the slowing-down frequency. The geometric relationships between the relevant variables are illustrated in Figure~\ref{fig:spherical_diagram}.

\section{Bosch-Hale parameterization of DT-$\alpha$ reaction}
\label{app:bosch_hale_parameterization}
The Bosch-Hale parameterization for the $DT-\alpha$ fusion Maxwellian averaged reactivity\cite{bosch_hale_1992_nucl_fusion_reactivity_legend_paper} is given as:
\begin{flalign}
    \label{eq:bh_parameterization}
    &
    \left< \sigma v\right> \left(T_{DT};\boldsymbol{\xi}\right)
    =
    \left\{
    \begin{array}{ccc}
        C_1 \theta \sqrt{\psi / \left( m_r c^2 T^3_{DT}  \right)} e^{-3\psi} & \text{if} & T_{DT} \in \left[0.2,100\right] \text{keV} \\
        0 & \text{otherwise} & 
    \end{array}
    \right\}.
    &
\end{flalign}
Here, $\boldsymbol{\xi} = \left\{ C_1, C_2, C_3, C_4, C_5, C_6, C_7, B_G, m_r c^2 \right\}$ is the parameter vector,
\begin{flalign}
    \label{eq:theta_for_bh_param}
    &
    \theta 
    = 
    \frac{T_{DT}}{1 - \frac{T_{DT} \left( C_2 + T_{DT} \left( C_4 + T_{DT} C_6 \right) \right)}
    {1 + T_{DT} \left(C_3 + T_{DT} \left( C_5 + T_{DT} C_7 \right) \right)}},
    &
\end{flalign}
\begin{flalign}
    \label{eq:psi_bs_param}
    &
    \psi = \left( \frac{B^2_G} {4 \theta} \right)^{1/3},
    &
\end{flalign}
$B_G = 34.3827~\sqrt{\text{keV}}$, $m_r c^2 = 1124656~\text{keV}$, $C_1 = 1.17302\times10^{-9}$, $C_2 = 1.51361\times10^{-2}$, $C_3 = 7.51886\times10^{-2}$, $C_4 = 4.60643\times10^{-3}$, $C_5 = 1.35 \times10^{-2}$, $C_6 = -1.06750\times10^{-4}$, and $C_7 = 1.366\times10^{-5}$. $T_{DT}$ must be provided in \unit{\keV} and the resulting reactivity will be in \unit{\cm^3\per\s}. For a detailed overview and the physical insights on the parameterization, we refer the reader to the original paper\cite{bosch_hale_1992_nucl_fusion_reactivity_legend_paper}.
%
%
%
%
%
\section*{Declaration of LLM usage in the writing process}
During the preparation of this work the authors used ChatGPT and Claude in order to improve the readability of the manuscript. After using this tool/service, the authors reviewed and edited the content as needed and take full responsibility for the content of the publication.
%
%
\bibliography{references}

\end{document}